\documentclass[12pt,a4paper]{article}
\usepackage{appendix}
\usepackage{algorithm}
\usepackage{algpseudocode}
\usepackage{geometry}
\usepackage{fancyhdr}
\usepackage[english]{babel}
\usepackage{cite}

\usepackage[utf8]{inputenc}
\usepackage[T1]{fontenc}
\RequirePackage{amsmath,amssymb}
\RequirePackage{mathrsfs}
\RequirePackage{amsfonts,bm}
\RequirePackage{dsfont}
\RequirePackage{braket}
\RequirePackage{tabularx}
\usepackage{multirow}
\RequirePackage{nicefrac}
\RequirePackage{graphicx}
\RequirePackage{booktabs}
\RequirePackage{colortbl}
\RequirePackage{xcolor}
\RequirePackage[symbol]{footmisc}
\usepackage{mathtools}
\usepackage{comment}
\usepackage{blkarray}
\usepackage{float}
\usepackage{rotating}
\usepackage{hhline}
\usepackage{tikz}
\usepackage{multirow}
\usepackage[section]{placeins}
\usepackage{cite}

\usepackage{amsmath}

\usepackage{graphicx}
\usepackage[colorlinks=true, allcolors=blue]{hyperref}
\def\beqn{\begin{eqnarray}}
\def\eeqn{\end{eqnarray}}

\usepackage{empheq}
\usepackage[most]{tcolorbox}
\usepackage{blkarray}
\newtcbox{\mymath}[1][]{%
    nobeforeafter, math upper, tcbox raise base,
    enhanced, colframe=blue!30!black,
    colback=blue!30, boxrule=1pt,
    #1}

\newcommand{\CC}[2]{C\genfrac[]{0pt}{}{#1}{#2}}

\newcommand{\ba}{\begin{eqnarray}}
\newcommand{\ea}{\end{eqnarray}}

\newcommand{\Sv}{\bm{S}}

\numberwithin{equation}{section}

\DeclareMathSymbol{\mg}{\mathrel}{symbols}{"1D}

\newcommand{\beq}{\begin{equation}}
\newcommand{\eeq}{\end{equation}}
\newcommand{\barr}{\begin{array}}
\newcommand{\earr}{\end{array}}

\newcounter{oldcounter}

\newcommand{\Bga}{{\boldsymbol \alpha}}
\newcommand{\Bgb}{{\boldsymbol \beta}}

\begin{document}
\begin{titlepage}
\samepage{
\setcounter{page}{1}
\rightline{July 2024}

\vfill
\begin{center}
  {\Large \bf{
      Vacuum Energy in Non--Supersymmetric\\ \medskip
    Quasi--Realistic Heterotic--String Vacua\\ \bigskip
    with Fixed Moduli
 }}

\vspace{1cm}
\vfill

{\large Eman Basaad $^{1}$\footnote{E-mail address: e.basaad@liverpool.ac.uk}, \large Luke A. Detraux $^{1}$\footnote{E-mail address: ldetraux@liverpool.ac.uk}, \\ \medskip
Alonzo R. Diaz Avalos$^{1}$\footnote{E-mail address: a.diaz-avalos@liverpool.ac.uk}, 
Alon E. Faraggi$^{1}$\footnote{E-mail address: alon.faraggi@liverpool.ac.uk},
 and Benjamin Percival$^{1,2}$\footnote{E-mail address: b.percival@mmu.ac.uk}}

\vspace{1cm}

{\it $^{1}$ Dept.\ of Mathematical Sciences, University of Liverpool, Liverpool
L69 7ZL, UK\\}
\vspace{.08in}
{\it $^{2}$ Dept. of Natural Sciences, Manchester Metropolitan University,  M15 6BH, UK \\}

\vspace{.025in}
\end{center}

\vfill

\begin{abstract}
Recently, Baykara, Tarazi and Vafa \cite{Baykara:2024vss, baykara2024new}
discussed the existence of quasicrystalline
string vacua that contain a single neutral moduli, the dilaton, and studied
compactifications of the non--supersymmetric $SO(16)\times SO(16)$
heterotic--string on these spaces.
We discuss a specific class
of quasi--realistic string vacua with similar properties that has been known
since the late eighties and analyse the vacuum energy in several
non--supersymmetric examples that correspond to compactifications of tachyon
free ten dimensional vacua as well as compactifications of tachyonic ten
dimensional vacua. Our analysis uses the Free Fermionic Formalism
of the heterotic-string in four dimensions and employs asymmetric
boundary conditions that project all the geometrical moduli
by Generalised GSO projections. This methodology produces models with both positive and negative spacetime potential at one--loop.
\end{abstract}
\smallskip}

\end{titlepage}
\newpage

\tableofcontents

\section{Introduction}
\label{sec:Intro}

String theory is the most developed contemporary framework
to explore the systhesis of the gauge and gravitational interactions.
Perturbative string theory predicts that a specific number of degrees of
freedom, beyond those that are observed in present--day experiments, are
required for its consistency. In some guise, some of these degrees
of freedom may be interpreted as extra spacetime dimensions, which
are compactified on an internal manifold.
The vast richness of the space of possibilities hinders the path
toward extracting the configuration which may correspond to our physical
world. However, the observed particle and cosmological data provide
strong constraints on the construction of viable models. 
The string vacua typically contain fields whose Vaccum Expectation Value (VEV) 
determine the characteristics of the internal manifolds and in turn fix
the phenomenological properties of the string models. Generic string
vacua contain a large number of such fields. However, string vacua that
are relevant for our physical world should contain few of those, 
if any at all. 

The question of the existence 
of stable De Sitter string vacua has generated substantial interest in 
string phenomenology over the past two decades. String vacua with positive vacuum energy 
exist in abundance \cite{FRPositiveCC, faraggi2020towards, Faraggi:2020wld}. 
The vital question is their stability. Typically, 
this question is investigated in an effective field theory limit of
the string vacua, although some progress has recently been made scanning the string landscape \cite{AFMPDterm, detraux2024StV, fraiman2023non}. 
Whether such effective field theory limits have a realisation in
string theory is an open question. However, closed string theory provides 
alternative routes to stabilise, or fix, the vacua. The independence 
of the left-- and right--moving solutions allows for their asymmetric 
treatment. This, for example, enables the construction of the 
heterotic--string \cite{heterotic}
in which the left--moving sector is fermionic, whereas 
the right--moving sector is bosonic. One can similarly assign asymmetric
actions on the degrees of freedom that correspond to the six dimensions 
of the compactified manifold.
This facilitates fixing of some or all of the internal dimensions
at fixed points in the moduli space. It implies that the associated 
neutral moduli fields, which allow the continuous deformations of the 
internal radii, are projected out from the physical spectrum. 

Phenomenological string models that reproduce the main characteristics
of the Standard Model, like the existence of three generations 
charged under a viable gauge group, {\it i.e.} one that may be reduced
to the Standard Model gauge group, were constructed since the late
1980s. A particular class that produces a rich space of quasi--realistic, 
three generation models is the class of heterotic--string models in the 
free fermionic formulation 
\cite{fsu5, fny,alr, slm, lrs, acfkr, su62, slmclass, lrsclass, cnonsusyPS}, 
which correspond to $\mathbb{Z}_2\times \mathbb{Z}_2$ toroidal orbifold compactifications at 
enhanced symmetry points in the 
moduli space \cite{z2z21}.
Many of the appealing phenomenological properties of the free fermionic
models are rooted in the underlying $\mathbb{Z}_2\times \mathbb{Z}_2$ orbifold 
structure \cite{z2z21}. 
In particular, as we discuss below in detail, this formulation facilitates the projection
of all the geometrical moduli, which imposes that the internal space is completely fixed. 
The projection of the geometrical moduli is generated by the utilisation
of asymmetric boundary conditions for the worldsheet fermions that correspond
to the internal compactified dimensions. 
However, the projection of all the geometrical moduli is achieved only 
in some special cases. 

In the fermionic worldsheet constructions, the marginal operators that 
generate the moduli deformations correspond to worldsheet Thirring interactions 
among the worldsheet fermions 
\cite{BaggerThirring, ChangKumar, LNY}. These worldsheet Thirring 
interactions correspond to massless fields in the string spectrum, 
which are the moduli fields. The allowed worldsheet Thirring interactions,
and the corresponding moduli fields, must be invariant under the Generalised
GSO (GGSO) projections. These GGSO projections are induced by the boundary condition basis vectors
that define the string models. For specific assignments, the worldsheet
Thirring interactions are forbidden, and the corresponding moduli fields are
projected out from the spectrum. In very special cases, 
all of the worldsheet Thirring interactions are forbidden and, therefore, 
all of the moduli fields are projected out. In those cases, 
all the geometrical coordinates are fixed at specific values 
in the moduli space. 
Furthermore, in the special cases that we discuss here, 
the projection of the moduli fields is obtained in tandem 
with the reduction of the number of chiral generations to three. 

It is important to emphasise that, while all the geometrical moduli
can be fixed in the models discussed in this paper, the dilaton
field remains unfixed at the perturbative level. To fix the dilaton field requires some nonperturbative 
effect, such as the racetrack mechanism \cite{racetrack, DixonRT}. This can be implemented 
in the vacua that we discuss here as they contain multiple hidden sector 
factors with varying number of matter states. The racetrack 
mechanism is implemented in the effective field theory limit 
and therefore will not be discussed further here. 
We note that discussions
of moduli fixing in the effective field theory limit of string 
compactifications date back to the early days of string phenomenology
\cite{rossetal}. 
The fixing of the geometrical moduli by the assignment of asymmetric 
boundary conditions can operate in supersymmetric vacua as well as
in non-supersymmetric string models. In the case of models with $N=1$ 
spacetime supersymmetry, the vacuum energy is identically zero and it 
has a finite value in non--supersymmetric models that can be either 
positive or negative, depending on the GGSO phase matrix. That is to say that in these non-supersymmetric models, there is no inherent necessity that the one--loop potential be positive or negative, and it can be manipulated by refining the GGSO matrix.

Motivated by the recent interest in non-supersymemtric heterotic models \cite{alonmirian, mavroudi, ashfaque2016non, Baykara:2024vss, angelantonj2024non, font2020exploring, parra2024non, nakajima2023new}, in this paper we calculate the 
vacuum energy of such models, constructed in the Free Fermionic Formalism. The models we focus on, derived from both tachyonic and tachyon--free ten dimensional vacua, have previously been shown to have favourable phenomenological properties \cite{faraggi2020stable, ashfaque2016non}, and have all geometric moduli fixed. We build on these models and construct examples in which the vacuum energy is positive and negative.
 In Section \ref{sec:FFF}, we review the Free Fermionic Formalism (FFF). Following this, Section \ref{sec:MF} gives a brief review of moduli fields in free fermionic models. We give a description of the partition function and potential under this formalism in Section \ref{sec:PF} and \ref{sec:Pot}. In Section \ref{sec:2} we review examples of non--supersymmetric models from previous works \cite{ashfaque2016non,faraggi2020stable}, calculating their potential, and adapting them further to find models with positive and negative values of the potential. Finally, we draw our conclusions and look towards future work in Section \ref{sec:con}.

\section{Free Fermionic Formalism}
\label{sec:FFF}
In this section we present an overview of the Free Fermionic Formalism (FFF) originally formalised within refs. \cite{antoniadis19884d, KLT,Antoniadis:4-dimensional} and recently reviewed in  \cite{Florakis:2024ubz}. In the FFF of the heterotic string in four dimensions, all the additional degrees of freedom ($18$ left-moving and $44$ right-moving) that are required to cancel the conformal anomaly are represented as free fermions propagating on the string worldsheet. In the light-cone gauge, the (worldsheet) supersymmetric left-moving sector includes the two transverse spacetime fermionic coordinates $\psi^{\mu}$ and 18 internal worldsheet real fermions. In the right-moving bosonic sector, the additional fermions are often represented as 12 real fermions, relating to the compactified manifold, and 16 complex fermions. The worldsheet fermions can propagate around the non-contracting loops of the torus and in doing so can therefore pick up a phase 
 
\begin{equation}
f \rightarrow -e^{i\pi\alpha(f)} f, \quad \alpha(f) \in (-1, +1].
\end{equation}
where $\alpha(f)$ is our boundary condition for the fermion $f$. Real boundary conditions are represented by $\alpha(f) \in \{0,1\}$, corresponding to Neveu-Schwarz or Ramond boundary conditions. In the following models it also becomes necessary to utilise complex boundary conditions, such that $\alpha(f) \in \{\frac{1}{2},-\frac{1}{2}\}$.
The phases between sectors are given as a GGSO phase matrix with elements, $\CC{\bm{v}_{i}}{\bm{v}_{j}}$, from which the Hilbert space is constructed. 

The construction of quasi-realistic free fermionic models involves the selection of specific basis vectors of boundary conditions. The general construction process follows two main steps. The first step of this process involves looking at the NAHE--set \cite{nahe}, a set of five basis vectors used to construct $SO(10)$ vacua. These vectors, $\bm{v}_{i}$, are $\left\{\mathds{1}, \Sv, \mathbf{b}_{1}, \mathbf{b}_{2}, \mathbf{b}_{3}\right\}$ defined as:
  \begin{equation}
\begin{aligned}
\label{eq:BV1}
\mathds{1} &\equiv  \{\psi^{\mu}, \chi^{1,...,6}, 
y^{1,...,6}, w^{1,...,6} ~| ~ \bar y^{1,...,6},\bar w^{1,...,6},\bar \eta^{1,2,3}, \bar \psi^{1,2,3,4,5}, \bar \phi^{1,...,8} \} \\
\Sv &\equiv \{\psi^{\mu}, \chi^{1,2}, \chi^{3,4},\chi^{5,6} \} \\
\mathbf{\mathbf{b}_{1}} &\equiv  \{\psi^{\mu}, \chi^{1,2}, y^{3,4,5,6} ~|~ \bar y^{3,4,5,6},\bar \eta^{1}, \bar \psi^{1,2,3,4,5} \} \\
\mathbf{\mathbf{b}_{2}} &\equiv \{\psi^{\mu}, \chi^{3,4}, y^{1,2}, w^{5,6} ~|~ \bar y^{1,2},\bar w^{5,6},\bar \eta^{2}, \bar \psi^{1,2,3,4,5} \} \\
\mathbf{\mathbf{b}_{3}} &\equiv \{\psi^{\mu}, \chi^{5,6}, w^{1,2,3,4} ~|~ \bar {w}^{1,2,3,4},\bar \eta^{3}, \bar \psi^{1,2,3,4,5} \} \text{.}\\
\end{aligned}
\end{equation}

In the second step of the construction, additional basis vectors 
are introduced to reduce the number of generations to three and break the four dimensional gauge group. A general additional basis vector can be defined as\\
\begin{equation}
\mathbf{b}_i=\left\{\alpha(\psi^{\mu}),...,\alpha(\omega^6) \mid \alpha(\bar{y}^1),...,\alpha(\bar \phi^8)\right\}  \text{ , }  
\end{equation}
where in general the labeling $\bm{b}_{4,5,6}$ indicate vectors that do not break the $SO(10)$ symmetry and $ \bm\alpha, \bm{\beta}, \bm\gamma$ indictating those that do. For example, $SO(10)$ is broken by the boundary conditions of $\bar{\psi}^{1, \ldots, 5}$ in $\bm\alpha, \bm\beta, \bm\gamma$, which can lead to $SU(5) \times U(1)$, $SO(6) \times SO(4)$ or $SU(3) \times SU(2) \times U(1)^2$ gauge groups. Each model we analyse in this paper follows this structure, with $\bm\alpha, \bm\beta, \bm\gamma$ varying to achieve different gauge groups and fulfil various phenomenological
conditions.
%
The partition function is then given by the sum of all possible sectors modulated by the GGSO matrix elements, as described in Section \ref{sec:PF}.
\section{Moduli Fields in Free Fermionic Models}
\label{sec:MF}

The phenomenological free fermionic heterotic--string models correspond to toroidal $\mathbb{Z}_2\times \mathbb{Z}_2$
orbifolds at special points in the moduli space. This correspondence is
discussed in detail in the literature \cite{z2z25, modulifix, FRtranslation} and elaborate dictionaries exist 
that facilitate translating the vacua from one representation to the other. 
In four dimensions the models are described in terms of two dimensional 
conformal and superconformal field theories with central charges $C_R=22$
and $C_L=9$, respectively. Deformations from the free fermionic point in the
moduli space are incorporated by worldsheet Thirring interactions between
the worldsheet fermions that are compatible with the conformal and modular 
invariance contraints. Untwisted moduli fields in the massless string
spectrum are in one--to--one correspondence with the coefficients of the allowed 
Thirring interactions. 

The exactly marginal operators associated with
untwisted moduli fields in symmetric orbifold models have the form
$\partial X^I{\bar\partial} X^J$, where $X^I$, $I=1,\cdots,6$, are
the coordinates of the six--torus $T^6$. 
The untwisted moduli fields admit a geometrical interpretation and 
appear as the couplings of the exactly marginal operators
in the non--linear sigma model action. In
the construction of the current algebra from chiral bosons, 
the operator $i\partial X^I$ is a $U(1)$ generator of the 
Cartan sub--algebra. In the fermionic
formalism,   $i\partial X^I_L\sim y^I\omega^I$ and 
$i\partial X^I_R\sim {\bar y}^I{\bar\omega}^I$, and
the exactly marginal operators are given by Abelian
Thirring operators of the form 
$J_L^i(z){\bar J}_R^j({\bar z})$,
where $J_L^i(z)$, ${\bar J}_R^j({\bar z})$
are some left-- and right--moving $U(1)$ currents 
in terms of worldsheet fermions. 
The untwisted moduli fields are the coefficients of the
Abelian Thirring interactions, which are invariant under the 
GGSO projections generated by the basis vectors in a given string model. 
The two dimensional action of the Thirring interactions takes the form 
\beq
S~=~\int d^2z h_{ij}(X) J_L^i(z) {\bar J}_R^j({\bar z})~\sim 
\int d^2z h_{ij} y^i\omega^i \bar{y}^j{\bar{\omega}}^j,
\label{2daction}
\eeq
where $J_L^i (i=1,\cdots ,6)$ are the left--moving chiral currents of
$U(1)^6$ 
and ${\bar J}^j_R (j=1,\cdots,22)$, are 
the right--moving chiral currents of $U(1)^{22}$. 

The models that
we consider here are NAHE \cite{nahe}
and ${\overline{\rm NAHE}}$ 
\cite{faraggi2020towards, faraggi2020stable} based models, 
where the 
NAHE--set is given by the set of five basis vectors described in Section \ref{sec:FFF}, 
$\left\{\mathds{1}, \Sv, \mathbf{b}_{1}, \mathbf{b}_{2}, \mathbf{b}_{3}\right\}$, and the 
$\overline{\rm NAHE}$--set is obtained by 
$\Sv\rightarrow \bm{\tilde{S}}$ map \cite{spwsp, faraggi2020towards}:
\begin{equation}
    \label{eq:S} \Sv = \{\psi^{\mu}, \chi^{1,2}, \chi^{3,4},\chi^{5,6}  \} \rightarrow \bm{\tilde{S}} = \{\psi^{\mu}, \chi^{1,2}, \chi^{3,4},\chi^{5,6} ~|~\bar{\phi}^{3,4,5,6}\} \text{ .}
\end{equation}
The Thirring interactions that are left invariant by the 
NAHE-- and $\overline{\rm NAHE}$--set are 
$$
J_L^{1,2}{\bar J}_R^{1,2}~~~~~~~~;~~~~~~~~%
J_L^{3,4}{\bar J}_R^{3,4}~~~~~~~~;~~~~~~~~%
J_L^{5,6}{\bar J}_R^{5,6}$$%
~~~~~~~~~~~~~~~~~~~
\beq
y^{1,2}\omega^{1,2}{\bar y}^{1,2}{\bar\omega}^{1,2}~~~~~;~~~~~%
y^{3,4}\omega^{3,4}{\bar y}^{3,4}{\bar\omega}^{3,4}~~~;~~~%
y^{5,6}\omega^{5,6}{\bar y}^{5,6}{\bar\omega}^{5,6}\text{ . } 
\label{thirringterms}
\eeq
 These set of untwisted 
moduli are present in all symmetric $\mathbb{Z}_2\times \mathbb{Z}_2$
orbifold models and correspond to the set of untwisted fields 
in these models. The corresponding scalar untwisted moduli 
fields from the Neveu--Schwarz sector are
\beq \label{hij}
h_{ij}=\ket{\chi^i}_L\otimes \ket{\bar{y}^j \bar{w}^j}_R=
\begin{cases} 
(i,j=1,2)\\
(i,j=3,4)\\
(i,j=5,6)
\end{cases}.
\eeq 
From these we can form the complex and K\"ahler structure moduli of the 
$\mathbb{Z}_2\times \mathbb{Z}_2$ orbifold that are given by 
\cite{LNY, modulifix}, 
\begin{align}
    \begin{split}
        T_1&=\frac{1}{\sqrt{2}}(H_1^{(1)}-iH_2^{(1)})=\frac{1}{\sqrt{2}}\ket{\chi^1+i\chi^2}_L\otimes \ket{\bar{y}^1 \bar{w}^1-i\bar{y}^2 \bar{w}^2}_R\\
        U_1&=\frac{1}{\sqrt{2}}(H_1^{(1)}+iH_2^{(1)})=\frac{1}{\sqrt{2}}\ket{\chi^1+i\chi^2}_L\otimes \ket{\bar{y}^1 \bar{w}^1+i\bar{y}^2 \bar{w}^2}_R
    \end{split}
\end{align}
and similarly for $T_{2,3}$ and $U_{2,3}$. The three complex structure 
and three K\"ahler structure moduli are present in all symmetric $\mathbb{Z}_2\times \mathbb{Z}_2$
orbifold compactifications. 

In the FFF we can 
assign asymmetric boundary conditions for the set of internal fermions
$\{y, \omega~|~{\bar y},{\bar\omega} \}$ that correspond to
the six left-- and right--moving fermionised coordinates. 
We remark that while the identification of the fermionised 
coordinates is fixed on the left--side by the super--current
constraint, there is some arbitrariness on the bosonic side, as discussed in ref. \cite{FGNP}. 
In the quasi--realistic free fermionic models the
symmetric versus asymmetric assignment is made in the basis vectors that extend 
the NAHE--set, which are constructed to reduce the number of generations to three and break the NAHE--based $SO(10)$ symmetry to one of its subgroups. Additional properties of the models, like the existence of untwisted electroweak Higgs doublets in the massless string spectrum and the existence of a leading Top Quark Yukawa coupling, 
depend on the assignment of symmetric 
versus asymmetric boundary conditions \cite{dtsm1994, yukawa}. 
An example of a three generation model with 
$SU(3)\times U(1)\times SU(2)^2$ unbroken $SO(10)$ subgroup
is given by
\beqn
 &\begin{tabular}{c|c|ccc|c|ccc|c}
 ~ & $\psi^\mu$ & $\chi^{12}$ & $\chi^{34}$ & $\chi^{56}$ &
        $\bar{\psi}^{1,...,5} $ &
        $\bar{\eta}^1 $&
        $\bar{\eta}^2 $&
        $\bar{\eta}^3 $&
        $\bar{\phi}^{1,...,8} $ \\
\hline
\hline
 ${\bm{\alpha}}$  &  0 & 0&0&0 & 1~1~1~0~0 & 0 & 0 & 0 &1~1~1~1~0~0~0~0 \\
 $\bm{\beta}$   &  0 & 0&0&0 & 1~1~1~0~0 & 0 & 0 & 0 &1~1~1~1~0~0~0~0 \\
 $\bm{\gamma}$  &  0 & 0&0&0 &
${\frac{1}{2}}$~${\frac{1}{2}}$~${\frac{1}{2}}$~0~0&${\frac{1}{2}}$&${\frac{1}{2}}$&${\frac{1}{2}}$
&0~${\frac{1}{2}}$~${\frac{1}{2}}$~${\frac{1}{2}}$~${\frac{1}{2}}$~${\frac{1}{2}}$~${\frac{1}{2}}$~0 \\
\end{tabular}
   \nonumber\\
   ~  &  ~ \nonumber\\
   ~  &  ~ \nonumber\\
     &\begin{tabular}{c|c|c|c}
 ~&   $y^3{y}^6$
      $y^4{\bar y}^4$
      $y^5{\bar y}^5$
      ${\bar y}^3{\bar y}^6$
  &   $y^1{\omega}^5$
      $y^2{\bar y}^2$
      $\omega^6{\bar\omega}^6$
      ${\bar y}^1{\bar\omega}^5$
  &   $\omega^2{\omega}^4$
      $\omega^1{\bar\omega}^1$
      $\omega^3{\bar\omega}^3$
      ${\bar\omega}^2{\bar\omega}^4$ \\
\hline
\hline
$\bm{\alpha}$& 1 ~~~ 1 ~~~ 1 ~~~ 0  & 1 ~~~ 1 ~~~ 1 ~~~ 0  & 1 ~~~ 1 ~~~ 1 ~~~ 0 \\
$\bm{\beta}$ & 0 ~~~ 1 ~~~ 0 ~~~ 1  & 0 ~~~ 1 ~~~ 0 ~~~ 1  & 1 ~~~ 0 ~~~ 0 ~~~ 0 \\
$\bm{\gamma}$& 0 ~~~ 0 ~~~ 1 ~~~ 1  & 1 ~~~ 0 ~~~ 0 ~~~ 0  & 0 ~~~ 1 ~~~ 0 ~~~ 1 \\
\end{tabular}
\label{model3}
\eeqn
This model gives rise 
to one type of (level-matched) tachyon producing sectors with
\beq
(\alpha_L^2,\alpha_R^2) ~ = ~ (2,6)  ~~~\text{and}~~~
                                 N_R ~=~ 0\label{ar6nr0}
\eeq
with the set of GGSO phases given by
\begin{equation}C \begin{blockarray}{c}
\begin{block}{[c]}
    \bm{v}_{i} \\
    \bm{v}_{j} \\
\end{block}
\end{blockarray} =  \text{  } 
{\bordermatrix{
          &{\mathds{1}}& \Sv & &{\mathbf{b}_{1}}&{\mathbf{b}_{2}}&{\mathbf{b}_{3}}& &{\bm\alpha}&{\bm\beta}&{\bm\gamma}\cr
       {\mathds{1}}&~~1&~~1 & & -1   &  -1 & -1  & & ~~1     & ~~1   & ~~i   \cr
             \Sv&~~1&~~1 & &~~1   & ~~1 &~~1  & & ~~1     & ~~1   &  -1   \cr
	      &   &    & &      &     &     & &         &       &       \cr
       {  \mathbf{b}_{1}}& -1& -1 & & -1   &  -1 & -1  & &  -1     &  -1   & ~~i   \cr
       {  \mathbf{b}_{2}}& -1& -1 & & -1   &  -1 & -1  & &  -1     &  -1   & ~~i   \cr
       {  \mathbf{b}_{3}}& -1& -1 & & -1   &  -1 & -1  & &  -1     & ~~1   & ~~i   \cr
	         &   &    & &      &     &     & &         &       &       \cr
      {\bm\alpha}&~~1&~~1 & &~~1   & ~~1 &~~1  & & ~~1     & ~~1   & ~~1   \cr
       {\bm\beta}&~~1&~~1 & & -1   &  -1 &~~1  & &  -1     &  -1   &  -1   \cr
      {\bm\gamma}&~~1& -1 & &~~1   &  -1 &~~1  & &  -1     &  -1   & ~~1   \cr}}.
\label{phasesmodel3}
\end{equation}

The full massless spectrum of this model together with the cubic level
superpotential
was presented in \cite{faraggi2020stable}.
It can easily be checked that all the terms in eq. (\ref{thirringterms})
are not invariant under the GGSO projections induced by the basis 
vectors in eq. (\ref{model3}), irrespective of the GGSO phases in eq. 
(\ref{phasesmodel3}). That is to say that the corresponding scalar fields are projected, and so in this model all the geometrical moduli
are fixed. Furthermore, the model does not contain any entirely
neutral fields aside from the dilaton. The model contains, like
many other models in this class, three untwisted states
that are neutral under the entire 
four dimensional gauge group. 
These
are obtained by acting on the NS vacuum with the 
oscillators 
$\chi_{12}{\bar \omega}^3{\bar \omega}^6|0\rangle$, 
$\chi_{34}{\bar \omega}^1{\bar y}^5|0\rangle$, 
$\chi_{56}{\bar y}^2{\bar y}^4|0\rangle$. 
Such states are ubiquitous in the free fermionic models. 
However, as seen from eq. (\ref{thirringterms}),
they do not correspond to moduli fields. They correspond to charged states
that become neutral due to the truncation of the rank
of the four dimensional gauge group
and carry discrete gauge charges. They arise because the free fermionic models 
are constructed at the enhanced symmetry point in the Narain moduli space. 
At the level of the extended NAHE--set \cite{gmh}, the right--moving world sheet fermions
$\{{\bar y},{\bar\omega}\}$ give rise to an enhanced $SO(4)^3$ gauge symmetry, 
corresponding to the $\{{\bar y}^{3\cdots 6}\};\{ {\bar y}^{1,2}, {\bar\omega}^{5,6}\}
\{{\bar\omega}^{1,\cdots, 4}\}$ groups of right--moving real worldsheet fermions, 
that are periodic in the sectors $b_1$, $b_2$ and $b_3$, respectively. One can
then combine pairs of these real worldsheet fermions to form the Cartan 
subalgebra and there is some freedom in the choice of these pairs, 
corresponding to the permutation symmetry of the right--moving 
real worldsheet fermions \cite{FGNP}. The completely neutral states in the model
of table \ref{model3} are then charged states. They become neutral states 
because of the reduction of the rank in the model that break the Cartan generators 
under which they are charged. We can see, however, that they do not correspond 
to geometrical moduli, which is our main interest here. This conclusion is borne
out by analysing the moduli in a bosonic interpretation of the model \cite{FGNP},
and observing that whatever combination of right--moving real fermions is taken,
the Thirring interactions and the corresponding moduli fields are 
always forbidden and projected out. This analysis confirms that in this
model all the geometrical moduli are fixed. We emphasise, and as emphasised 
in \cite{modulifix, FGNP}, that this is not generically the case and is particular to the
class of models to which the model in table \ref{model3} belongs. Specifically, to the 
pairings of the real right--moving worldsheet fermions. 
In general, they give rise to non--vanishing
terms in the cubic level superpotential and therefore generically will become
massive in supersymmetric preserving vacua along 
$F$-- and $D$--flat directions \cite{nonrenoterms}.
We note that the model still contains numerous charged fields and a fully dynamical 
analysis of the vacuum is yet to be performed. The space 
of charged fields can further be constrained by using a combination of symmetric
and asymmetric boundary conditions with respect to the set of internal fermions
$\{y,\omega\vert {\bar y},{\bar\omega}\}^{1,\cdots,6}$, as {\it e.g.} in the model
of Table \ref{stringmodel}
\beqn
 &\begin{tabular}{c|c|ccc|c|ccc|c}
 ~ & $\psi^\mu$ & $\chi^{12}$ & $\chi^{34}$ & $\chi^{56}$ &
        $\bar{\psi}^{1,...,5} $ &
        $\bar{\eta}^1 $&
        $\bar{\eta}^2 $&
        $\bar{\eta}^3 $&
        $\bar{\phi}^{1,...,8} $ \\
\hline
\hline
  $\bm{\alpha}$  & ~0 &~0&0&0 &~1~1~1~0~0 &~1 &~0 &~0 &~1~1~0~0~0~0~0~0 \\
  $\bm{\beta}$   & ~0 &~0&0&0 &~1~1~1~0~0 &~0 &~1 &~0 &~0~0~1~1~0~0~0~0 \\
  $\bm{\gamma}$  & ~0 &~0&0&0 &
		${\frac{1}{2}}$~${\frac{1}{2}}$~${\frac{1}{2}}$~${\frac{1}{2}}$~${\frac{1}{2}}$
	      & ${\frac{1}{2}}$ & ${\frac{1}{2}}$ & ${\frac{1}{2}}$ &
               ~0~0~0~0~$\frac{1}{2}$~$\frac{1}{2}$~${\frac{1}{2}}$~${\frac{1}{2}}$ \\
\end{tabular}
   \nonumber\\
   ~  &  ~ \nonumber\\
   ~  &  ~ \nonumber\\
     &\begin{tabular}{c|c|c|c}
 ~&   $y^3{y}^6$
      $y^4{\bar y}^4$
      $y^5{\bar y}^5$
      ${\bar y}^3{\bar y}^6$
  &   $y^1{\omega}^5$
      $y^2{\bar y}^2$
      $\omega^6{\bar\omega}^6$
      ${\bar y}^1{\bar\omega}^5$
  &   $\omega^2{\omega}^4$
      $\omega^1{\bar\omega}^1$
      $\omega^3{\bar\omega}^3$
      ${\bar\omega}^2{\bar\omega}^4$ \\
\hline
\hline
$\bm\alpha$ &~1 ~~~~0 ~~~~0 ~~~~1  &~0 ~~~~0 ~~~~1 ~~~~1  &~0 ~~~~0 ~~~~1 ~~~~1 \\
$\bm\beta$  &~0 ~~~~0 ~~~~1 ~~~~1  &~1 ~~~~0 ~~~~0 ~~~~1  &~0 ~~~~1 ~~~~0 ~~~~1 \\
$\bm\gamma$ &~0 ~~~~1 ~~~~0 ~~~~0  &~0 ~~~~1 ~~~~0 ~~~~0  &~1 ~~~~0 ~~~~0 ~~~~0 \\
\end{tabular}
\label{stringmodel}
\eeqn
As can be checked from eq. (\ref{thirringterms}), all the Thirring interaction terms
are not invariant under the GGSO projections defined by the basis vectors 
in eq. (\ref{stringmodel}). However, from the boundary conditions in eq. 
(\ref{stringmodel}) we note for example that the boundary conditions 
with respect to the set of internal fermions $\{{\bar y}^{3,\cdots,6}\}$
is symmetric in $\alpha$ but asymmetric in $\beta$ and the same
is the case with respect to the set of fermions 
$\{{\bar y}^{1,2},{\bar\omega}^{5,6}\}$, whereas both are asymmetric 
with respect to the set of fermions $\{{\bar\omega}^{1,\cdots,4}\}$. 
The basis vectors $\alpha$ and $\beta$
both break the $SO(10)$ symmetry of the NAHE--set to the $SO(6)\times SO(4)$
subgroup. The consequence of assigning a mixture of symmetric and asymmetric 
boundary conditions is the reduction in the number of charged fields in the 
model \cite{cleaver2008quasi, Cleaver:2011ir}. 

To summarise this section we note that the boundary condition basis vectors 
in eqs. (\ref{model3}) and (\ref{stringmodel}) forces the pairing 
of the pairs of real fermions 
$y^1\omega^5,~\omega^2\omega^4~{\rm and}~y^3y^6$ into complex fermions. 
It entails that none of the worldsheet Thirring interactions in eq. 
(\ref{thirringterms}) are allowed by these boundary conditions and that
all of the associated moduli fields in eq. (\ref{hij}) are 
projected out by the GGSO projections. Hence, the internal space in
models that utilise this pairing is completely fixed. Ref. \cite{FGNP}
provided a bosonic interpretation of this construction. 

\section{Partition Function}
\label{sec:PF}
In the FFF, the partition function can be calculated in the following modular invariant form:
\begin{equation}
    \label{eq:Partfunctferm}
\mathbf{Z} = Z_{B} \sum_{St} C\begin{bmatrix} \bm\alpha \\ \bm\beta \end{bmatrix} \prod Z\begin{bmatrix} \alpha(f) \\ \beta(f) \end{bmatrix} \text{ , }
\end{equation}
\begin{equation}
Z\begin{bmatrix} 1 \\ 1 \end{bmatrix} = \sqrt{\frac{\vartheta_{1}}{\eta}} \text{ , }Z\begin{bmatrix} 1 \\ 0 \end{bmatrix} = \sqrt{\frac{\vartheta_{2}}{\eta}} \text{ , }Z\begin{bmatrix} 0 \\ 0 \end{bmatrix} = \sqrt{\frac{\vartheta_{3}}{\eta}} \text{ , }Z\begin{bmatrix} 0 \\ 1 \end{bmatrix} = \sqrt{\frac{\vartheta_{4}}{\eta}} \text{ , }
\end{equation}
where $Z_{B}$ describes the bosonic contribuition and is given by
\begin{equation}
    Z_{B} = \frac{1}{\tau_{2}} \frac{1}{\eta^{2}\bar{\eta}^{2}} \text{ . }
\end{equation}
    The sectors are labelled by $\Bga$ and $\Bgb$; the sum is over the sectors and the product is over the fermions in each sector, with complex fermions contributing to the product twice; $\tau_{2}$ is the imaginary component of the modular parameter. Definitions of $\vartheta$ and $\eta$/$\bar{\eta}$ in terms of $\tau$/$\bar{\tau}$ can be found in Appendix A of \cite{faraggi2020towards}.

 For non-supersymmetric models, it is more useful to express the partition function more overtly as a polynomial in terms of the `nome' $q \equiv e^{2i\pi(\tau_{1}+i\tau_{2})}$ and $p \equiv \bar{q} \equiv e^{-2i\pi(\tau_{1}-i\tau_{2})}$. The general form of this polynomial is 
\begin{equation}
    Z = \sum_{m,n} \frac{a_{mn}}{\tau_{2}}q^{m}p^n \text{ , }
\end{equation}
 where the additional factor of $\tau_{2}$ is omitted from the polynomial in the following examples, and reintroduced during integration. In this form the coefficients $a_{mn}$ correspond to the difference between the number of bosonic and fermionic states, $N_{b} - N_{f}$, at mass level $(m, n)$. Moreover, divergent terms can be identified easily in the polynomial. This idea is discussed further in Section \ref{sec:Pot}.
\section{Potential}
\label{sec:Pot}
Once the partition function has been found and expressed as a polynomial, the spacetime potential can be found from integrating this over the fundamental domain of the torus. 

\begin{equation}
    \label{eq:Pot}
    \begin{split}
V_{1-loop} &= - \frac{1}{2} \frac{\mathcal{M}^{4}}{(2\pi)^{4}} \int_{\mathcal{F}} \frac{d^{2}\tau}{\tau^{2}_{2}} Z(\tau, \bar\tau ; T^{(i)}, U^{(i)})\\
&= - \frac{1}{2} \frac{\mathcal{M}^{4}}{(2\pi)^{4}} \int_{\mathcal{F}} \frac{d^{2}\tau}{\tau^{3}_{2}} \sum a_{mn}q^{m}\bar{q}^n\\
&= \sum a_{mn} I_{mn} \text{ , }
    \end{split}
\end{equation}
where $\frac{d^{2}\tau}{\tau^{2}_{2}}$
is the modular invariant measure and the fundamental domain, $\mathcal{F}$, is defined as:
\begin{equation}
    \mathcal{F} = \mathcal{F}_{1} + \mathcal{F}_{2}
\end{equation}
\begin{equation}
    \mathcal{F}_{1} = \{ \tau \in \mathbb{C} \text{ }| \quad \tau_{2} \geq 1 \quad \land \quad |\tau_{1}| < \frac{1}{2} \}
\end{equation}
\begin{equation}
    \mathcal{F}_{2} = \{ \tau \in \mathbb{C} \text{ }| \quad |\tau|^{2} > 1 \quad \land \quad \tau_{2} < 1\quad \land \quad |\tau_{1}| < \frac{1}{2} \} \text{ . }
\end{equation}
Importantly, it can be shown that the integral over $\mathcal{F}_{2}$ will always be finite, however the conditions for finiteness over $\mathcal{F}_{1}$ are as follows \cite{faraggi2020towards}:
\begin{equation}
\label{eq:inf}
I_{mn} = 
\begin{cases} 
    \infty & \text{if } m+n < 0 \text{ and } m-n \not\in \mathbb{Z} \setminus \{0\} \\
    \text{Finite} & \text{otherwise.}
\end{cases}
\end{equation}
Because of this, both level-matched and non level-matched tachyonic states can lead to divergences and destabilise the vacuum. For NAHE-- and ${\overline{\rm NAHE}}$--based models with only real boundary conditions, a finite vacuum energy can be achieved by simply projecting out the level-matched tachyons. We believe that this can be generalised to imaginary boundary conditions also, and it is indeed the case in the models we consider. 

\section{Analysis of Models}
\label{sec:2}
\subsection{$S$--Models}
\label{subsec:Model1}
As we have discussed, there are multiple ways to break supersymmetry in this formalism. The first way we will consider is through GGSO projections acting on a model derived from a tachyon free $SO(10)$ vacuum. This was considered in \cite{ashfaque2016non}, and we adopt the same basis vectors and GGSO matrix below. 
We remark that classifying the supersymmetry breaking as explicit or spontaneous requires analysis of the dependence of the model on the geometric moduli, in order to establish 
whether supersymmetry is restored on the boundary of the moduli space, in which case it is 
classified as "spontaneous--breaking". There is not other simple criteria that informs us
whether the breaking is spontaneous or explicit. 
As there is no dependence of the vacuum energy here on any moduli, there is no evidence for either conclusion. Ref \cite{ashfaque2016non} discusses how some sectors maintain their supersymmetric structure, whilst in other sectors this is not the case
\begin{equation}
\begin{aligned}
\label{eq:BV}
\mathds{1} &= \bm{v}_{1} =  \{\psi^{\mu}, \chi^{1,...,6}, y^{1,...,6}, w^{1,...,6} ~|~ \bar y^{1,...,6},\bar w^{1,...,6},\bar \eta^{1,2,3}, \bar \psi^{1,2,3,4,5}, \bar \phi^{1,...,8} \} \\
\Sv &= \bm{v}_2 = \{\psi^{\mu}, \chi^{1,2}, \chi^{3,4},\chi^{5,6} \} \\
\mathbf{b}_{1} &= \bm{v}_{3} =   \{\psi^{\mu}, \chi^{1,2}, y^{3,4,5,6} ~|~ \bar y^{3,4,5,6},\bar \eta^{1}, \bar \psi^{1,2,3,4,5} \} \\
\mathbf{b}_{2} &= \bm{v}_{4} = \{\psi^{\mu}, \chi^{3,4}, y^{1,2}, w^{5,6} ~|~ \bar y^{1,2},\bar w^{5,6},\bar \eta^{2}, \bar \psi^{1,2,3,4,5} \} \\
\mathbf{b}_{3} &= \bm{v}_{5} = \{\psi^{\mu}, \chi^{5,6}, w^{1,2,3,4} ~|~\bar w^{1,2,3,4},\bar \eta^{3}, \bar \psi^{1,2,3,4,5} \} \\
\bm\alpha &= \bm{v}_{6} =\{y^{1,2,3,4,5,6}, w^{1,2,3,4,5,6}~|~ \bar{y}^{2,4,5}, \bar{w}^{1,3,6}, \bar{\psi}^{1,2,3}, \bar{\phi}^{1,2,3,4} \} \\
\bm\beta &= \bm{v}_{7} = \{y^{2,4},{w}^{2,4} ~|~ \bar{y}^{1,2,3,4, 6}, \bar{w}^{5},  \bar{\psi}^{1,2,3}, \bar{\phi}^{1,2,3,4} \} \\
\bm\gamma &= \bm{v}_{8} = \{{y}^{1,5} w^{1,5} ~|~ \bar{y}^{3,5,6},\bar{w}^{1,2,4}, \bar{\psi}^{1,2,3} = \frac{1}{2}, \bar{\eta}^{1,2,3} = \frac{1}{2}, \bar{\phi}^{2,3,4,5,6,7} = \frac{1}{2} \} \\
\end{aligned}
\end{equation}
It was shown in Section \ref{sec:MF} that these basis vectors fix the geometric moduli and project the associated scalar fields. Depending on the choice of GGSO phases, one can build supersymmetric and non-supersymmetric models. Below we present the GGSO matrix previously used to construct a non-supersymmetric model:
\begin{equation}C \begin{blockarray}{c}
\begin{block}{[c]}
    \bm{v}_{i} \\
    \bm{v}_{j} \\
\end{block}
\end{blockarray} =  \text{  } 
{\bordermatrix{
          &{\mathds{1}}& \Sv & &{\mathbf{b}_{1}}&{\mathbf{b}_{2}}&{\mathbf{b}_{3}}& &{\bm\alpha}&{\bm\beta}&{\bm\gamma}\cr
       {\mathds{1}}&~~1&~~1 & & -1   &  -1 & -1  & & ~~1     & ~~1   & ~~i   \cr
             \Sv&~~1&~~1 & &~~1   & ~~1 &~~1  & & ~~1     & ~~1   &  -1   \cr
	      &   &    & &      &     &     & &         &       &       \cr
       {  \mathbf{b}_{1}}& -1& -1 & & -1   &  -1 & -1  & &  -1     &  -1   & ~~i   \cr
       {  \mathbf{b}_{2}}& -1& -1 & & -1   &  -1 & -1  & &  -1     &  -1   & ~~i   \cr
       {  \mathbf{b}_{3}}& -1& -1 & & -1   &  -1 & -1  & &  -1     & ~~1   & ~~i   \cr
	         &   &    & &      &     &     & &         &       &       \cr
      {\bm\alpha}&~~1&~~1 & &~~1   & ~~1 &~~1  & & ~~1     & ~~1   & ~~1   \cr
       {\bm\beta}&~~1&~~1 & & -1   &  -1 &-1  & &  -1     &  -1   &  -1   \cr
      {\bm\gamma}&~~1& -1 & &~~1   &  -1 &~~1  & &  -1     &  -1   & ~~1   \cr}}.
\end{equation}
The full spectrum of this model is given in Appendix A. 
The gauge group of the model is: 
\begin{equation}
    SU(3)_{C} \times U(1)_{C} \times SU(2)_{L} \times SU(2)_{R} \times \prod^{6}_{i=1} U_{i} \times SU(3)_{H_{1}} \times SU(3)_{H_{2}} \times \prod^{10}_{j=7} U_{j}
\end{equation}

The survival of supersymmetry in this model is dependent on $\CC{\Sv}{\bm\alpha}$ and $\CC{\Sv}{\bm\beta}$, and can be restored through the following modification:
\begin{equation}
    C\begin{bmatrix} \Sv \\ \bm\alpha  \\ \end{bmatrix}  \rightarrow -1 \text{, and  }
C\begin{bmatrix} \Sv \\ \bm\beta  \\ \end{bmatrix} \rightarrow -1
\end{equation}
The supersymmetric case of course gives a vanishing partition function when calculated, whereas the non-supersymmetric model returns the following partition function:

\begin{equation}
\begin{split}
    Z =& ~56+\frac{2}{p} + \frac{56q}{p} -  \frac{16q^{\frac{1}{2}}}{p^{\frac{1}{2}}}+288 q+2048 p^{\frac{1}{4}} q^{\frac{1}{4}}+\frac{128 p^{\frac{1}{2}}}{q^{\frac{1}{2}}}-27648 p^{\frac{1}{2}} q^{\frac{1}{2}}\\&+\frac{8704 p^{\frac{3}{4}}}{q^{\frac{1}{4}}}-410624 p^{\frac{3}{4}} q^{\frac{3}{4}}+138048 p+1494784 p q ... \text{ , }
    \end{split}
\end{equation}
where we define $q = e^{2i\pi\tau}$ and $p = \bar{q} = e^{-2i\pi\bar{\tau}} $. 
We give the partition function here to $\mathcal{O}(1)$ in $p$ and $q$ as higher order terms give diminishing corrections to the potential, and our aim is simply to determine if the potential we find is finite and to give an example of a model with a positive cosmological constant.


Integrating the partition function, as defined in Section \ref{sec:Pot}, we find the model returns a positive Cosmological Constant, corresponding to a De Sitter vacuum
\begin{equation}
    \Lambda = 0.00499799 \mathcal{M}_{s}^{4} \text{ . }
\end{equation}

 In pursuit of additional examples, we modify the previous GGSO matrix in the following way
 \begin{equation}
    C\begin{bmatrix} \bm\beta \\ \bm{b}_{3}  \\ \end{bmatrix} \rightarrow 1 \text{, and  }
C\begin{bmatrix}  \bm{b}_{3}  \\ \bm\beta \\ \end{bmatrix} \rightarrow -1\text{, }
\end{equation}
giving the following GGSO matrix:
\begin{equation}C \begin{blockarray}{c}
\begin{block}{[c]}
    \bm{v}_{i} \\
    \bm{v}_{j} \\
\end{block}
\end{blockarray} =  \text{  } 
{\bordermatrix{
          &{\mathds{1}}& \Sv & &{\mathbf{b}_{1}}&{\mathbf{b}_{2}}&{\mathbf{b}_{3}}& &{\bm\alpha}&{\bm\beta}&{\bm\gamma}\cr
       {\mathds{1}}&~~1&~~1 & & -1   &  -1 & -1  & & ~~1     & ~~1   & ~~i   \cr
             \Sv&~~1&~~1 & &~~1   & ~~1 &~~1  & & ~~1     & ~~1   &  -1   \cr
	      &   &    & &      &     &     & &         &       &       \cr
       {  \mathbf{b}_{1}}& -1& -1 & & -1   &  -1 & -1  & &  -1     &  -1   & ~~i   \cr
       {  \mathbf{b}_{2}}& -1& -1 & & -1   &  -1 & -1  & &  -1     &  -1   & ~~i   \cr
       {  \mathbf{b}_{3}}& -1& -1 & & -1   &  -1 & -1  & &  -1     & -1   & ~~i   \cr
	         &   &    & &      &     &     & &         &       &       \cr
      {\bm\alpha}&~~1&~~1 & &~~1   & ~~1 &~~1  & & ~~1     & ~~1   & ~~1   \cr
       {\bm\beta}&~~1&~~1 & & -1   &  -1 &~~1  & &  -1     &  -1   &  -1   \cr
      {\bm\gamma}&~~1& -1 & &~~1   &  -1 &~~1  & &  -1     &  -1   & ~~1   \cr}}.
\end{equation}

Whilst the coefficients of the partition function differ, the cosmological constant remains positive and finite: 
\begin{equation}
    \begin{split}
        Z =& ~8 + \frac{2}{p}+ \frac{56 q}{p}+\frac{32 q^{\frac{1}{2}}}{p^{\frac{1}{2}}}+\frac{512 q^{\frac{3}{4}}}{p^{\frac{1}{4}}} -3360 q+1024 p^{\frac{1}{4}} q^{\frac{1}{4}} +\frac{192 p^{\frac{1}{2}}}{q^{\frac{1}{2}}}-24576 p^{\frac{1}{2}} q^{\frac{1}{2}}\\ & +\frac{8704 p^{\frac{3}{4}}}{q^{\frac{1}{4}}}-346112 p^{\frac{3}{4}} q^{\frac{3}{4}}+138240 p+1427456 p q... \text{ , }
    \end{split}
\end{equation}

\begin{equation}
    \Lambda = 0.0174667 \mathcal{M}_{s}^{4} \text{ . }
\end{equation}

\subsection{$\tilde{S}$--Models}
\label{subsec:Model2}
The second route to non-supersymetric models is through explicit supersymmetry breaking at the $SO(10)$ level, via the $\Sv \rightarrow \bm{\tilde{S}}$ map \cite{spwsp, faraggi2020towards}. The following supersymmetric basis set and GGSO matrix was defined in \cite{cleaver2008quasi} and  later adapted to the $\tilde{S}$--model in \cite{faraggi2020stable}:
\begin{equation}
\begin{aligned}
\label{eq:BV2}
\mathds{1}&= \bm{v}_{1}=  \{\psi^{\mu}, \chi^{1,...,6}, y^{1,...,6}, w^{1,...,6} ~|~ \bar y^{1,...,6}, \bar w^{1,...,6}, \bar \eta^{1,2,3}, \bar \psi^{1,2,3,4,5}, \bar \phi^{1,...,8} \} \\
\Sv&= \bm{v}_2 = \{\psi^{\mu}, \chi^{1,2}, \chi^{3,4}, \chi^{5,6} \} \\
\mathbf{b}_{1}&= \bm{v}_{3} = \{\psi^{\mu}, \chi^{1,2}, y^{3,4,5,6} ~|~ \bar y^{3,4,5,6}, \bar \eta^{1}, \bar \psi^{1,2,3,4,5} \} \\
\mathbf{b}_{2}& = \bm{v}_{4} = \{\psi^{\mu}, \chi^{3,4}, y^{1,2}, w^{5,6} ~|~ \bar y^{1,2}, \bar w^{5,6}, \bar \eta^{2}, \bar \psi^{1,2,3,4,5} \} \\
\mathbf{b}_{3}& = \bm{v}_{5}= \{\psi^{\mu}, \chi^{5,6}, w^{1,2,3,4} ~|~ \bar w^{1,2,3,4}, \bar \eta^{3}, \bar \psi^{1,2,3,4,5} \} \\
\bm\alpha & = \bm{v}_{6} = \{y^{3}y^{6}, \bar{y}^{3}\bar{y}^{6}, w^{6}\bar{w}^{6}, \bar{y}^{1}\bar{w}^{5}, w^{3}\bar{w}^{3}, \bar{w}^{2}\bar{w}^{4}, \bar{\psi}^{1,2,3}, \bar{\eta}^{1}, \bar{\phi}^{1,2} \} \\
\bm{\beta}& = \bm{v}_{7} = \{y^{5}\bar{y}^{5}, \bar{y}^{3}\bar{y}^{6}, y^{1}{w}^{5}, \bar{y}^{1}\bar{w}^{5}, w^{1}\bar{w}^{1}, \bar{w}^{2}\bar{w}^{4}, \bar{\psi}^{1,2,3}, \bar{\eta}^{2}, \bar{\phi}^{3,4} \} \\
\bm{\gamma}& = \bm{v}_{8} = \{{y}^{4}\bar{y}^{4}, y^{2}\bar{y}^{2}, {w}^{2}{w}^{4}, \bar{\psi}^{1,2,3,4,5} = \frac{1}{2}, \bar{\eta}^{1,2,3} = \frac{1}{2}, \bar{\phi}^{5,6,7,8} = \frac{1}{2} \} \\
\end{aligned}
\end{equation}
\begin{equation}C \begin{blockarray}{c}
\begin{block}{[c]}
    \bm{v}_{i} \\
    \bm{v}_{j} \\
\end{block}
\end{blockarray} =  \text{  } 
{\bordermatrix{
          &{\mathds{1}}& \Sv & &{\mathbf{b}_{1}}&{\mathbf{b}_{2}}&{\mathbf{b}_{3}}& &{\bm\alpha}&{\bm\beta}&{\bm\gamma}\cr
       {\mathds{1}}&~~1&~~1 & & -1   &  -1 & -1  & & -1     & -1   & ~~i   \cr
             \Sv&~~1&~~1 & &~~1   & ~~1 &~~1  & & -1     & -1   &  -1   \cr
	      &   &    & &      &     &     & &         &       &       \cr
       {  \mathbf{b}_{1}}& -1& -1 & & -1   &  -1 & -1  & &  -1     &  -1   & ~~i   \cr
       {  \mathbf{b}_{2}}& -1& -1 & & -1   &  -1 & -1  & &  -1     &  ~~1   & ~~i   \cr
       {  \mathbf{b}_{3}}& -1& -1 & & -1   &  -1 & -1  & &  ~~1     & -1   & ~~1   \cr
	         &   &    & &      &     &     & &         &       &       \cr
      {\bm\alpha}&-1&-1 & &-1   & -1 &~~1  & & ~~1     & ~~1   & ~~1   \cr
       {\bm\beta}&-1&-1 & & -1   &  ~~1 &-1  & &  -1     &  ~~1   &  ~~1   \cr
      {\bm\gamma}&-1& -1 & &~~1   &  ~~1 &-1  & &  -1     &  -1   & -i   \cr}}.
\end{equation}
\\
Following the pattern of Section \ref{subsec:Model1}, we begin with a stable supersymmetric model with vanishing partition function and cosmological constant. 

In \cite{faraggi2020stable}, the following modifications are discussed:
\begin{equation}
  \Sv \rightarrow \bm{\tilde{S}} = \{\psi^{\mu}, \chi^{1,2}, \chi^{3,4},\chi^{5,6} ~|~ \bar{\phi}^{3,4,5,6} \}  \text{ , }
\end{equation}
\begin{equation}
  C\begin{bmatrix}  \bm{\tilde{S}}  \\ \bm{\gamma} \\ \end{bmatrix} 
  \rightarrow i \text{ ,  } C\begin{bmatrix}  \bm{\beta}  \\ \bm{\tilde{S}} \\ \end{bmatrix} 
  \rightarrow 1  \text{. }
\end{equation}
We further develop the matrix in the following way:
\begin{equation}
C\begin{bmatrix} \bm{\tilde{S}} \\ \bm{\tilde{S}} \\ \end{bmatrix} \rightarrow -1\text{ ,  }
  C\begin{bmatrix} \bm\alpha \\ \bm\alpha \\ \end{bmatrix} 
  \rightarrow -1 \text{ ,  } C\begin{bmatrix} \bm\alpha \\ \mathds{1}  \\ \end{bmatrix} 
  \rightarrow 1  \text{ ,  } C\begin{bmatrix} \mathds{1} \\ \bm\alpha  \\ \end{bmatrix} 
  \rightarrow 1\text{.}
\end{equation}
For clarity and completeness, the resulting matrix is given below.
\begin{equation}C \begin{blockarray}{c}
\begin{block}{[c]}
    \bm{v}_{i} \\
    \bm{v}_{j} \\
\end{block}
\end{blockarray} =  \text{  } 
{\bordermatrix{
          &{\mathds{1}}& \bm{\tilde{S}} & &{\mathbf{b}_{1}}&{\mathbf{b}_{2}}&{\mathbf{b}_{3}}& &{\bm\alpha}&{\bm\beta}&{\bm\gamma}\cr
       {\mathds{1}}&~~1&~~1 & & -1   &  -1 & -1  & & ~~1     & -1   & ~~i   \cr
             \bm{\tilde{S}}&~~1&-1 & &~~1   & ~~1 &~~1  & & -1     & -1   &  ~~i   \cr
	      &   &    & &      &     &     & &         &       &       \cr
       {  \mathbf{b}_{1}}& -1& -1 & & -1   &  -1 & -1  & &  -1     &  -1   & ~~i   \cr
       {  \mathbf{b}_{2}}& -1& -1 & & -1   &  -1 & -1  & &  -1     &  ~~1   & ~~i   \cr
       {  \mathbf{b}_{3}}& -1& -1 & & -1   &  -1 & -1  & &  ~~1     & -1   & ~~1   \cr
	         &   &    & &      &     &     & &         &       &       \cr
      {\bm\alpha}&~~1&-1 & &-1   & -1 &~~1  & & -1     & ~~1   & ~~1   \cr
       {\bm\beta}&-1&~~1 & & -1   &  ~~1 &-1  & &  -1     &  ~~1   &  ~~1   \cr
      {\bm\gamma}&-1& -1 & &~~1   &  ~~1 &-1  & &  -1     &  -1   & -i   \cr}}.
      \label{eq:stilde1GGSO}
\end{equation}
The gauge group is now enhanced and is given by 
\begin{equation}
    SU(3)_{C} \times U(1)_{C} \times SU(2)_{L} \times U(1)_{L} \times \prod^{6}_{i=1} U_{i} \times SU(4) \times \prod^{4}_{j=1} SU(2)_{j} \times U(1) \text{ ,}
\end{equation}
where the last three terms are contributions from the hidden sector. The full spectrum is given in Appendix B. 
Applying the same formula, we find the partition function and vacuum energy to be the following:

\begin{equation}
    \begin{split}
Z =& ~~168+ \frac{2}{p}+ \frac{56 q}{p} -\frac{16q^{\frac{1}{2}}}{p^{\frac{1}{2}}}+\frac{64q^{\frac{3}{4}}}{p^{\frac{1}{4}}}-\frac{240q^{\frac{5}{8}}}{p^{\frac{3}{8}}} +864q+ 336q^{\frac{1}{8}}p^{\frac{1}{8}}
+1408q^{\frac{1}{4}}p^{\frac{1}{4}}\\
& + \frac{96p^{\frac{1}{2}}}{q^{\frac{1}{2}}}-27776q^{\frac{1}{2}}p^{\frac{1}{2}}-22320 q^{\frac{5}{8}}p^{\frac{5}{8}}+\frac{8256p^{\frac{3}{4}}}{q^{\frac{1}{4}}}
-395520q^{\frac{3}{4}}p^{\frac{3}{4}}\\& +142464p+1530368qp ...\text{ , }\\
    \end{split}
\end{equation}
\begin{equation}
    \Lambda = -0.0199 \mathcal{M}_{s}^{4}\text{ .}
\end{equation}
Here we see that at the one-loop level, we find a finite negative cosmological constant, providing a counter example to the suggestion in \cite{baykara2024new} that  ``more rigid” tachyon-free string theories always have positive cosmological constant.

\section{Discussion and Conclusion}
\label{sec:con}
In this paper we analyse non-supersymmetric 
heterotic--string models with all geometric moduli fixed, and 
calculated the vacuum energy in such models. This paper follows the work of Baykara, Tarazi and Vafa \cite{baykara2024new}, who recently presented models of similar properties, constructed using quasicrystaline orbifolds. Angelantonj, Florakis, Leone and Perugini \cite{angelantonj2024non} have since also presented non-tachyonic, non-supersymmetric heterotic vacua with this property.
We used the free fermionic formulation to construct the string vacua that correspond to $\mathbb{Z}_2\times \mathbb{Z}_2$ toroidal orbifold compactifications
at special points in the moduli spaces. 
The internal spaces that we utilised in our investigations were used
since the late eighties in the construction of phenomenological 
three generation string models and led, for example, to a prediction of
the top quark mass several years prior to its experimental observation
\cite{tqmp}.

This method is well established and builds on the NAHE--set with three additional basis vectors with asymmetric boundary conditions. These basis vectors project the scalars associated with the geometric moduli and fixes the internal space, such that only the dilaton is unfixed. The quasi-realistic models we consider here have previously been shown to conform to the following phenomenological conditions: projection of level matched tachyons; three generations of chiral fermions; and contain a Higgs doublet. The starting points to achieve these phenomenological models are quite different, and have been achieved through compactification of a tachyon free ten dimensional vacuum in $S$--Models, and through compactification of a tachyonic ten dimensional vacuum in $\tilde{S}$--Models. 
We have shown in Sections \ref{subsec:Model1} and \ref{subsec:Model2} that both routes to non-supersymmetric models can produce vacua with finite potential, and that the value of this one--loop potential can be positive or negative, relating to De Sitter and anti--De Sitter spaces.

However, there is a number of phenomenological issues with the models we have presented here. Consider the $S$--Model, which starts from a tachyon free, supersymmetric vacuum and has a finite, positive cosmological constant when supersymmetry is broken through GGSO projection. In Ref. \cite{ashfaque2016non}, it is noted, " ...in this model the untwisted Higgs bi-doublets,
which couple at leading order to the twisted sector states,
are projected out and consequently the leading mass term
which is identified with the top mass is absent". Therefore work remains to be done to find a model which adheres to the TQMC fertility condition \cite{rizos2014top} without compromising the stability of the vacuum potential. We have no reason to assume these two conditions are mutually exclusive. Similarly, there is no reason to assume the vacuum potential will always be positive in $S$-Models, though we only give positive examples above.
 Turning our attention to the $\tilde{S}$--Model of Ref, \cite{faraggi2020stable}, which arises from a tachyonic ten dimensional vacuum. 
 By modifying the GGSO phase matrix, we show it is possible to construct models with finite, negative cosmological constant, and provide the new spectrum for this model. 
 As in the $S$--Model, there is no reason \textit{a priori} that an $\tilde{S}$--Model cannot be found that meets the fertility conditions and has a finite potential, which may be positive or negative. It is our understanding that the infrastructure to find such models already exists in the methodology described above. Incorporating the 'fertility methodology' as discussed in e.g. Ref \cite{cnonsusyPS}, we conjecture that stable, fertile models with all geometric moduli fixed can be found in abundance. 
 
 As a word of caution we remark that the question of the stability of non--supersymmetric
 string vacua should be further examined. In the first place, 
 the question of stability at higher orders remains open, as emphasised in refs. \cite{ashfaque2016non} and \cite{faraggi2020stable}, 
higher order terms may cause instabilities in the potentials. Similarly, a full analysis of
the potential of the string vacua is yet to be performed, {\it e.g.} hidden sector 
condensates may stabilise the dilaton but may also destablise the vacuum. 
Furthermore, while it is found quite generally that all the geometrical moduli are 
projected out in the models studied here, the role of the three untwisted states
that are neutral under the 
four dimensional gauge group,
$\chi_{12}{\bar \omega}^3{\bar \omega}^6|0\rangle$, 
$\chi_{34}{\bar \omega}^1{\bar y}^5|0\rangle$, 
$\chi_{56}{\bar y}^2{\bar y}^4|0\rangle$, 
ought to be further understood. 
These questions are not unique to the models we have presented here and are ubiquitous across both supersymmetric and non-supersymmetric models. However we view this work as a natural extension to the phenomenological analysis of these models in previous papers.

\section*{Acknowledgements}
This research was supported in part by grant NSF PHY-2309135 to the Kavli Institute for Theoretical Physics (KITP). 
The work of ARDA is supported in part by EPSRC grant EP/T517975/1.
The work of LD is supported in part by STFC grant, and in part by LIV.INNO .
AEF would like to thank the Kavli Institute for Theoretical Physics, 
the CERN theory division, and the Weizmann Institute for hospitality. 


\appendix
\section{Spectrum of Model in Section 6.1}
The notation for the table is the following: The first column describes if the states correspond to spacetime bosons or spacetime fermions and specifically for $b_i$ the type of particle. The second column is the name of the sector.
The third column gives the dimensionality of the states under $SU(3)_C \times
SU(2)_L \times SU(2)_R$ and the fourth the charges of the observable $U(1)$s.
Columns 5 and 6 describe the hidden sector. The only charges appearing in the
table that do not have a self--evident name are:
\begin{eqnarray}
Q_C&=&
Q_{\overline{\psi}^{1}}+Q_{\overline{\psi}^{2}}+Q_{\overline{\psi}^{3}}~,\nonumber\\
Q_8&=&
Q_{\overline{\phi}^{2}}+Q_{\overline{\phi}^{3}}+Q_{\overline{\phi}^{4}}~,\nonumber\\
Q_9&=&
Q_{\overline{\phi}^{5}}+Q_{\overline{\phi}^{6}}+Q_{\overline{\phi}^{7}}~.
\end{eqnarray}
To avoid writing fractional numbers all the charges in the table have been
multiplied by $4$. Finally, for every state the CPT conjugate is also
understood to be in the spectrum and has not been written explicitly.
\begin{table}[!ht]
\hspace*{-18 mm}
\begin{tabular}{|c|c|c|ccccccc|c|cccc|}
\hline
$F$&
SEC&$(C;L;R)$&$Q_{C}$&$Q_{\bar{\eta}^1}$&$Q_{\bar{\eta}^2}$&$Q_{\bar{\eta}^3}$&$Q_{\bar{y}^{3,6}}$&$Q_{\bar{y}^1\bar{w}^5}$&$Q_{\bar{w}^{2,4}}$&$SU(3)_{H_{1,2}}$&$Q_{\bar{\Phi}^1}$&$Q_{8}$&$Q_{9}$&$Q_{\bar{\Phi}^8}$\\
\hline
b&NS&$(1,1,1)$&\,\,$0$&\,\,$0$&\,\,$0$&\,\,$0$&\,\,$0$&\,\,$0$&\,\,$0$&$(1,1)$&\,\,0&\,\,0&\,\,0&\,\,0\\
&&$(1,1,1)$&\,\,$0$&\,\,$0$&\,\,$0$&\,\,$0$&\,\,$0$&\,\,$0$&\,\,$0$&$(1,1)$&\,\,0&\,\,0&\,\,0&\,\,0\\
&&$(1,1,1)$&\,\,$0$&\,\,$0$&\,\,$0$&\,\,$0$&\,\,$0$&\,\,$0$&\,\,$0$&$(1,1)$&\,\,0&\,\,0&\,\,0&\,\,0\\
&&$(1,1,1)$&\,\,$0$&-$4$&\,\,$4$&\,\,$0$&\,\,$0$&\,\,$0$&\,\,$0$&$(1,1)$&\,\,0&\,\,0&\,\,0&\,\,0\\
&&$(1,1,1)$&\,\,$0$&\,\,$4$&-$4$&\,\,$0$&\,\,$0$&\,\,$0$&\,\,$0$&$(1,1)$&\,\,0&\,\,0&\,\,0&\,\,0\\
&&$(1,1,1)$&\,\,$0$&\,\,$0$&-$4$&\,\,$4$&\,\,$0$&\,\,$0$&\,\,$0$&$(1,1)$&\,\,0&\,\,0&\,\,0&\,\,0\\
&&$(1,1,1)$&\,\,$0$&\,\,$0$&\,\,$4$&-$4$&\,\,$0$&\,\,$0$&\,\,$0$&$(1,1)$&\,\,0&\,\,0&\,\,0&\,\,0\\
&&$(1,1,1)$&\,\,$0$&-$4$&\,\,$0$&\,\,$4$&\,\,$0$&\,\,$0$&\,\,$0$&$(1,1)$&\,\,0&\,\,0&\,\,0&\,\,0\\
&&$(1,1,1)$&\,\,$0$&\,\,$4$&\,\,$0$&-$4$&\,\,$0$&\,\,$0$&\,\,$0$&$(1,1)$&\,\,0&\,\,0&\,\,0&\,\,0\\
\hline

\hline
\end{tabular}
\caption{The untwisted Neveu-Schwarz sector matter states and charges.}
\end{table}
\begin{table}
\hspace*{-18 mm}
\begin{tabular}{|c|c|c|ccccccc|c|cccc|}
\hline
$F$&
SEC&$(C;L;R)$&$Q_{C}$&$Q_{\bar{\eta}^1}$&$Q_{\bar{\eta}^2}$&$Q_{\bar{\eta}^3}$&$Q_{\bar{y}^{3,6}}$&$Q_{\bar{y}^1\bar{w}^5}$&$Q_{\bar{w}^{2,4}}$&$SU(3)_{H_{1,2}}$&$Q_{\bar{\Phi}^1}$&$Q_{8}$&$Q_{9}$&$Q_{\bar{\Phi}^8}$\\
\hline

f&${S}$&$(1,1,1)$&\,\,$0$&\,\,$0$&\,\,$0$&\,\,$0$&\,\,$0$&\,\,$0$&\,\,$0$&$(3,\overline{3})$&\,\,0&-$4$&\,\,$4$&\,\,0\\
&&$(1,1,1)$&\,\,$0$&\,\,$0$&\,\,$0$&\,\,$0$&\,\,$0$&\,\,$0$&\,\,$0$&$(3,\overline{3})$&\,\,0&\,\,4&-$4$&\,\,0\\
&&$(1,1,1)$&\,\,$0$&\,\,$0$&\,\,$0$&\,\,$0$&\,\,$0$&\,\,$0$&\,\,$0$&$(1,1)$&\,\,$4$&\,\,0&\,\,0&\,\,$4$\\
&&$(1,1,1)$&\,\,$0$&\,\,$0$&\,\,$0$&\,\,$0$&\,\,$0$&\,\,$0$&\,\,$0$&$(1,1)$&\,\,$4$&\,\,0&\,\,0&-$4$\\
&&$(1,1,1)$&\,\,$0$&\,\,$0$&\,\,$0$&\,\,$0$&\,\,$0$&\,\,$0$&\,\,$0$&$(1,1)$&-$4$&\,\,0&\,\,0&\,\,$4$\\
&&$(1,1,1)$&\,\,$0$&\,\,$0$&\,\,$0$&\,\,$0$&\,\,$0$&\,\,$0$&\,\,$0$&$(1,1)$&-$4$&\,\,0&\,\,0&-$4$\\

&&$(3,1,1)$       &-$4$&\,\,$4$&\,\,$0$&\,\,$0$&\,\,$0$&\,\,$0$&\,\,$0$&$(1,1)$&\,\,0&\,\,0&\,\,0&\,\,0\\
\hline
\end{tabular}
\caption{The untwisted $S$--sector matter states and charges.}
\end{table}

\newpage
\begin{center}
\begin{table}
\hspace*{-18 mm}
\begin{tabular}{|c|c|c|ccccccc|c|cccc|}
\hline
$F$&
SEC&$(C;L;R)$&$Q_{C}$&$Q_{\bar{\eta}^1}$&$Q_{\bar{\eta}^2}$&$Q_{\bar{\eta}^3}$&$Q_{\bar{y}^{3,6}}$&$Q_{\bar{y}^1\bar{w}^5}$&$Q_{\bar{w}^{2,4}}$&$SU(3)_{H_{1,2}}$&$Q_{\bar{\Phi}^1}$&$Q_{8}$&$Q_{9}$&$Q_{\bar{\Phi}^8}$\\
\hline
$Q_{L_{1}}$&${{b}}_{1}$&$(3,2,1)$&\,\,$2$&\,\,$2$&\,\,$0$&\,\,$0$&-$2$&\,\,$0$&\,\,$0$&$(1,1)$&0&0&0&0\\
$Q_{R_{1}}$&&$(\overline{3},1,2)$&-$2$&-$2$&\,\,$0$&\,\,$0$&-$2$&\,\,$0$&\,\,$0$&$(1,1)$&0&0&0&0 \\
$L_{L_{1}}$&&$(1,2,1)$&-$6$&\,\,$2$&\,\,$0$&\,\,$0$&-$2$&\,\,$0$&\,\,$0$&$(1,1)$&0&0&0&0\\
$L_{R_{1}}$&&$(1,1,2)$&\,\,$6$&-$2$&\,\,$0$&\,\,$0$&-$2$&\,\,$0$&\,\,$0$&$(1,1)$&0&0&0&0\\
\hline
b&${S}+{b_{1}}$&$(3,1,2)$&\,\,$2$&\,\,$2$&\,\,$0$&\,\,$0$&-$2$&\,\,$0$&\,\,$0$&$(1,1)$&$0$&$0$&$0$&$0$\\
&&$(\overline{3},2,1)$&-$2$&-$2$&\,\,$0$&\,\,$0$&-$2$&\,\,$0$&\,\,$0$&$(1,1)$&$0$&$0$&$0$&$0$\\
&&$(1,2,1)$&\,\,$6$&-$2$&\,\,$0$&\,\,$0$&-$2$&\,\,$0$&\,\,$0$&$(1,1)$&$0$&$0$&$0$&$0$\\
\hline
\end{tabular}
\caption{The observable matter sectors.}
\end{table}
\end{center}

\newpage
\begin{center}
\begin{table}[!ht]
\hspace*{-18 mm}
\begin{tabular}{|c|c|c|ccccccc|c|cccc|}
\hline
$F$&
SEC&$(C;L;R)$&$Q_{C}$&$Q_{\bar{\eta}^1}$&$Q_{\bar{\eta}^2}$&$Q_{\bar{\eta}^3}$&$Q_{\bar{y}^{3,6}}$&$Q_{\bar{y}^1\bar{w}^5}$&$Q_{\bar{w}^{2,4}}$&$SU(3)_{H_{1,2}}$&$Q_{\bar{\Phi}^1}$&$Q_{8}$&$Q_{9}$&$Q_{\bar{\Phi}^8}$\\
\hline
f&${S}+$&$(1,1,1)$&\,\,$0$&\,\,$2$&-$2$&\,\,$0$&\,\,$0$&\,\,$0$&\,\,$0$&$(1,1)$&\,\,$0$&\,\,$0$&\,\,$0$&\,\,$4$\\
&${b_{1}}+{b_{2}}$&$(1,1,1)$&\,\,$0$&\,\,$2$&-$2$&\,\,$0$&\,\,$0$&\,\,$0$&\,\,$0$&$(1,1)$&\,\,$0$&\,\,$0$&\,\,$0$&-$4$\\
&$+\alpha+\beta$&$(1,1,1)$&\,\,$0$&-$2$&\,\,$2$&\,\,$0$&\,\,$0$&\,\,$0$&\,\,$0$&$(1,1)$&\,\,$0$&\,\,$0$&\,\,$0$&\,\,$4$\\
&&$(1,1,1)$&\,\,$0$&-$2$&\,\,$2$&\,\,$0$&\,\,$0$&\,\,$0$&\,\,$0$&$(1,1)$&\,\,$0$&\,\,$0$&\,\,$0$&-$4$\\
&&$(1,1,1)$&\,\,$0$&\,\,$2$&\,\,$2$&\,\,$0$&\,\,$0$&\,\,$0$&\,\,$0$&$(3,1)$&\,\,$0$&\,\,$4$&\,\,$0$&\,\,$0$\\
&&$(1,1,1)$&\,\,$0$&-$2$&-$2$&\,\,$0$&\,\,$0$&\,\,$0$&\,\,$0$&$(3,1)$&\,\,$0$&\,\,$4$&\,\,$0$&\,\,$0$\\
&&$(1,1,1)$&\,\,$0$&\,\,$2$&\,\,$2$&\,\,$0$&\,\,$0$&\,\,$0$&\,\,$0$&$(\overline{3},1)$&\,\,$0$&-$4$&\,\,$0$&\,\,$0$\\
\hline
\end{tabular}
\caption{Vector-like $SO(10)$ singlet states.}
\end{table}
\end{center}

\newpage
\begin{center}
\begin{table}[!ht]
\hspace*{-18 mm}
\begin{tabular}{|c|c|c|ccccccc|c|cccc|}
\hline
$F$&
SEC&$(C;L;R)$&$Q_{C}$&$Q_{\bar{\eta}^1}$&$Q_{\bar{\eta}^2}$&$Q_{\bar{\eta}^3}$&$Q_{\bar{y}^{3,6}}$&$Q_{\bar{y}^1\bar{w}^5}$&$Q_{\bar{w}^{2,4}}$&$SU(3)_{H_{1,2}}$&$Q_{\bar{\Phi}^1}$&$Q_{8}$&$Q_{9}$&$Q_{\bar{\Phi}^8}$\\
\hline
f&${S}+$&$(1,1,1)$&\,\,$0$&-$2$&\,\,$0$&\,\,$0$&\,\,$0$&-$2$&-$2$&$(1,1)$&$0$&\,\,$6$&$0$&\,\,$2$\\
&${1}+{b_{1}}$&$(1,1,1)$&\,\,$0$&\,\,$2$&\,\,$0$&\,\,$0$&\,\,$0$&-$2$&-$2$&$(1,1)$&$0$&-$6$&$0$&-$2$\\
&$+\alpha+2\gamma$&$(1,1,1)$&\,\,$0$&\,\,$2$&\,\,$0$&\,\,$0$&\,\,$0$&-$2$&-$2$&$(3,1)$&$0$&\,\,$2$&$0$&-$2$\\
&&$(1,1,1)$&\,\,$0$&-$2$&\,\,$0$&\,\,$0$&\,\,$0$&-$2$&-$2$&$(\overline{3},1)$&$0$&-$2$&$0$&\,\,$2$\\
\hline
b&${1}+{b_{1}}$&$(1,1,1)$&\,\,$0$&-$2$&\,\,$0$&\,\,$0$&\,\,$0$&\,\,$2$&-$2$&$(1,1)$&\,\,$0$&\,\,$6$&\,\,$0$&\,\,$2$\\
&$+\alpha+2\gamma$&$(1,1,1)$&\,\,$0$&\,\,$2$&\,\,$0$&\,\,$0$&\,\,$0$&\,\,$2$&-$2$&$(1,1)$&\,\,$0$&-$6$&\,\,$0$&-$2$\\
&&$(1,1,1)$&\,\,$0$&\,\,$2$&\,\,$0$&\,\,$0$&\,\,$0$&\,\,$2$&-$2$&$(3,1)$&\,\,$0$&\,\,$2$&\,\,$0$&-$2$\\
\hline
\end{tabular}
\caption{Vector-like $SO(10)$ singlet states.}
\end{table}
\end{center}

\newpage
\begin{center}
\begin{table}[!ht]
\hspace*{-18 mm}
\begin{tabular}{|c|c|c|ccccccc|c|cccc|}
\hline
$F$&
SEC&$(C;L;R)$&$Q_{C}$&$Q_{\bar{\eta}^1}$&$Q_{\bar{\eta}^2}$&$Q_{\bar{\eta}^3}$&$Q_{\bar{y}^{3,6}}$&$Q_{\bar{y}^1\bar{w}^5}$&$Q_{\bar{w}^{2,4}}$&$SU(3)_{H_{1,2}}$&$Q_{\bar{\Phi}^1}$&$Q_{8}$&$Q_{9}$&$Q_{\bar{\Phi}^8}$\\
\hline
f&${S}+$&$(1,1,1)$&\,\,$0$&-$2$&\,\,$0$&\,\,$0$&\,\,$0$&-$2$&-$2$&$(1,1)$&-$2$&\,\,$0$&\,\,$6$&\,\,$0$\\
&${b_{2}}+{b_{3}}$&$(1,1,1)$&\,\,$0$&\,\,$2$&\,\,$0$&\,\,$0$&\,\,$0$&-$2$&-$2$&$(1,1)$&\,\,$2$&\,\,$0$&-$6$&\,\,$0$\\
&$+\alpha+2\gamma$&$(1,1,1)$&\,\,$0$&\,\,$2$&\,\,$0$&\,\,$0$&\,\,$0$&-$2$&-$2$&$(1,3)$&\,\,$2$&\,\,$0$&\,\,$2$&\,\,$0$\\
&&$(1,1,1)$&\,\,$0$&-$2$&\,\,$0$&\,\,$0$&\,\,$0$&-$2$&-$2$&$(1,\overline{3})$&-$2$&\,\,$0$&-$2$&\,\,$0$\\
\hline
b&${b_{2}}+{b_{3}}$&$(1,1,1)$&\,\,$0$&-$2$&\,\,$0$&\,\,$0$&\,\,$0$&\,\,$2$&-$2$&$(1,1)$&-$2$&\,\,$0$&\,\,$6$&\,\,$0$\\
&$+\alpha+2\gamma$&$(1,1,1)$&\,\,$0$&\,\,$2$&\,\,$0$&\,\,$0$&\,\,$0$&\,\,$2$&-$2$&$(1,1)$&\,\,$2$&$0$&-$6$&\,\,$0$\\
&&$(1,1,1)$&\,\,$0$&\,\,$2$&\,\,$0$&\,\,$0$&\,\,$0$&\,\,$2$&-$2$&$(1,3)$&\,\,$2$&$0$&\,\,$2$&\,\,$0$\\
\hline
\end{tabular}
\caption{Table 5 continued.}
\end{table}
\end{center}

\newpage
\begin{center}
\begin{table}[!ht]
\hspace*{-18 mm}
\begin{tabular}{|c|c|c|ccccccc|c|cccc|}
\hline
$F$&
SEC&$(C;L;R)$&$Q_{C}$&$Q_{\bar{\eta}^1}$&$Q_{\bar{\eta}^2}$&$Q_{\bar{\eta}^3}$&$Q_{\bar{y}^{3,6}}$&$Q_{\bar{y}^1\bar{w}^5}$&$Q_{\bar{w}^{2,4}}$&$SU(3)_{H_{1,2}}$&$Q_{\bar{\Phi}^1}$&$Q_{8}$&$Q_{9}$&$Q_{\bar{\Phi}^8}$\\
\hline
b&$\alpha\pm\gamma$&$(1,1,1)$&-$3$&\,\,$1$&\,\,$1$&\,\,$1$&-$2$&\,\,$0$&-$2$&$(1,1)$&\,\,$2$&-$3$&\,\,$3$&\,\,$0$\\
&&$(1,1,1)$&-$3$&\,\,$1$&\,\,$1$&\,\,$1$&\,\,$2$&\,\,$0$&\,\,$2$&$(1,1)$&\,\,$2$&-$3$&\,\,$3$&\,\,$0$\\
&&$(1,1,1)$&-$3$&\,\,$1$&\,\,$1$&\,\,$1$&\,\,$2$&\,\,$0$&\,\,$2$&$(1,1)$&\,\,$2$&-$3$&\,\,$3$&\,\,$0$\\
&&$(1,1,1)$&-$3$&\,\,$1$&\,\,$1$&\,\,$1$&-$2$&\,\,$0$&-$2$&$(1,1)$&\,\,$2$&-$3$&\,\,$3$&\,\,$0$\\
&&$(1,1,1)$&\,\,$3$&-$1$&-$1$&-$1$&-$2$&\,\,$0$&-$2$&$(1,1)$&-$2$&\,\,$3$&-$3$&\,\,$0$\\
&&$(1,1,1)$&\,\,$3$&-$1$&-$1$&-$1$&\,\,$2$&\,\,$0$&\,\,$2$&$(1,1)$&-$2$&\,\,$3$&-$3$&\,\,$0$\\
&&$(1,1,1)$&\,\,$3$&-$1$&-$1$&-$1$&\,\,$2$&\,\,$0$&\,\,$2$&$(1,1)$&-$2$&\,\,$3$&-$3$&\,\,$0$\\
\hline
\end{tabular}
\caption{All the massless sectors for which the ``would-be superpartners" are massive and do not form part of the massless spectrum. The ``would-be superpartners" arise from the sectors that are obtained by adding the basis vector $S$ to a given sector and are the  fermionic counterparts.}
\end{table}
\end{center}

\newpage
\begin{center}
\begin{table}
\hspace*{-18 mm}
\begin{tabular}{|c|c|c|ccccccc|c|cccc|}
\hline
$F$&
SEC&$(C;L;R)$&$Q_{C}$&$Q_{\bar{\eta}^1}$&$Q_{\bar{\eta}^2}$&$Q_{\bar{\eta}^3}$&$Q_{\bar{y}^{3,6}}$&$Q_{\bar{y}^1\bar{w}^5}$&$Q_{\bar{w}^{2,4}}$&$SU(3)_{H_{1,2}}$&$Q_{\bar{\Phi}^1}$&$Q_{8}$&$Q_{9}$&$Q_{\bar{\Phi}^8}$\\
\hline
f&${S}+$&$(1,1,1)$&-$3$&-$3$&-$1$&-$1$&\,\,$0$&\,\,$0$&\,\,$0$&$(1,1)$&-$2$&-$3$&\,\,$3$&\,\,$0$\\
&${b_{2}}+{b_{3}}$&$(1,1,1)$&-$3$&\,\,$1$&\,\,$3$&-$1$&\,\,$0$&\,\,$0$&\,\,$0$&$(1,1)$&-$2$&-$3$&\,\,$3$&\,\,$0$\\
&$+\beta\pm\gamma$&$(1,1,1)$&-$3$&\,\,$1$&-$1$&\,\,$3$&\,\,$0$&\,\,$0$&\,\,$0$&$(1,1)$&-$2$&-$3$&\,\,$3$&\,\,$0$\\
&&$(1,1,1)$&\,\,$3$&-$1$&\,\,$1$&\,\,$1$&\,\,$0$&\,\,$0$&\,\,$0$&$(1,\overline{3})$&-$2$&\,\,$3$&\,\,$1$&\,\,$0$\\
&&$(1,1,1)$&\,\,$3$&\,\,$3$&\,\,$1$&\,\,$1$&\,\,$0$&\,\,$0$&\,\,$0$&$(1,1)$&\,\,$2$&\,\,$3$&-$3$&\,\,$0$\\
&&$(1,1,1)$&\,\,$3$&-$1$&-$3$&\,\,$1$&\,\,$0$&\,\,$0$&\,\,$0$&$(1,1)$&\,\,$2$&\,\,$3$&-$3$&\,\,$0$\\
&&$(1,1,1)$&\,\,$3$&-$1$&\,\,$1$&-$3$&\,\,$0$&\,\,$0$&\,\,$0$&$(1,1)$&\,\,$2$&\,\,$3$&-$3$&\,\,$0$\\
\hline
\end{tabular}
\caption{Vector-like exotic states.}
\end{table}
\end{center}

\newpage
\begin{center}
\begin{table}[!ht]
\hspace*{-18 mm}
\begin{tabular}{|c|c|c|ccccccc|c|cccc|}
\hline
$F$&
SEC&$(C;L;R)$&$Q_{C}$&$Q_{\bar{\eta}^1}$&$Q_{\bar{\eta}^2}$&$Q_{\bar{\eta}^3}$&$Q_{\bar{y}^{3,6}}$&$Q_{\bar{y}^1\bar{w}^5}$&$Q_{\bar{w}^{2,4}}$&$SU(3)_{H_{1,2}}$&$Q_{\bar{\Phi}^1}$&$Q_{8}$&$Q_{9}$&$Q_{\bar{\Phi}^8}$\\
\hline
f&${S}+$&$(\overline{3},1,1)$&-$3$&-$1$&\,\,$1$&-$1$&\,\,$0$&\,\,$0$&\,\,$0$&$(1,1)$&\,\,$0$&\,\,$3$&-$3$&-$2$\\
&${1}+{b_{2}}$&$(1,1,1)$&\,\,$3$&-$1$&\,\,$1$&-$1$&\,\,$0$&\,\,$0$&\,\,$0$&$(1,\overline{3})$&\,\,$0$&\,\,$3$&\,\,$3$&\,\,$2$\\
&$+\alpha\pm\gamma$&$(3,1,1)$&\,\,$3$&\,\,$1$&-$1$&\,\,$1$&\,\,$0$&\,\,$0$&\,\,$0$&$(1,1)$&$\,\,0$&-$3$&$3$&\,\,$2$\\
&&$(1,1,1)$&-$3$&\,\,$1$&-$1$&\,\,$1$&\,\,$0$&\,\,$0$&\,\,$0$&$(1,3)$&\,\,$0$&-$3$&-$3$&-$2$\\
\hline
b&${1}+{b_{2}}$&$(1,1,1)$&-$3$&\,\,$3$&\,\,$1$&-$1$&\,\,$0$&\,\,$0$&\,\,$0$&$(1,1)$&\,\,$0$&\,\,$3$&-$3$&-$2$\\
&$+\alpha\pm\gamma$&$(1,1,1)$&-$3$&-$1$&-$3$&-$1$&\,\,$0$&\,\,$0$&\,\,$0$&$(1,1)$&\,\,$0$&\,\,$3$&-$3$&-$2$\\
&&$(1,1,1)$&-$3$&-$1$&\,\,$1$&\,\,$3$&\,\,$0$&\,\,$0$&\,\,$0$&$(1,1)$&\,\,$0$&\,\,$3$&-$3$&-$2$\\
\hline
\end{tabular}
\caption{Vector-like exotic states (continued).}
\end{table}
\end{center}

\newpage
\begin{center}
\begin{table}[!ht]
\hspace*{-18 mm}
\begin{tabular}{|c|c|c|ccccccc|c|cccc|}
\hline
$F$&
SEC&$(C;L;R)$&$Q_{C}$&$Q_{\bar{\eta}^1}$&$Q_{\bar{\eta}^2}$&$Q_{\bar{\eta}^3}$&$Q_{\bar{y}^{3,6}}$&$Q_{\bar{y}^1\bar{w}^5}$&$Q_{\bar{w}^{2,4}}$&$SU(3)_{H_{1,2}}$&$Q_{\bar{\Phi}^1}$&$Q_{8}$&$Q_{9}$&$Q_{\bar{\Phi}^8}$\\
\hline
f&${1}+{b_{2}}$&$(1,2,1)$&\,\,$0$&\,\,$0$&-$2$&-$2$&\,\,$2$&\,\,$0$&\,\,$0$&$(1,1)$&-$2$&\,\,$0$&\,\,$0$&\,\,$2$\\
&$+{{b_{3}}}+2\gamma$&$(1,2,1)$&\,\,$0$&\,\,$0$&-$2$&-$2$&-$2$&\,\,$0$&\,\,$0$&$(1,1)$&\,\,$2$&\,\,$0$&\,\,$0$&-$2$\\
&&$(1,1,2)$&\,\,$0$&\,\,$0$&\,\,$2$&\,\,$2$&\,\,$2$&\,\,$0$&\,\,$0$&$(1,1)$&\,\,$2$&\,\,$0$&\,\,$0$&-$2$\\
&&$(1,1,2)$&\,\,$0$&\,\,$0$&\,\,$2$&\,\,$2$&-$2$&\,\,$0$&\,\,$0$&$(1,1)$&-$2$&\,\,$0$&\,\,$0$&\,\,$2$\\
\hline
b&${S}+$&$(1,2,1)$&\,\,$0$&\,\,$0$&\,\,$2$&\,\,$2$&-$2$&\,\,$0$&\,\,$0$&$(1,1)$&-$2$&\,\,$0$&\,\,$0$&\,\,$2$\\
&${1}+{b_{2}}$&$(1,2,1)$&\,\,$0$&\,\,$0$&\,\,$2$&\,\,$2$&\,\,$2$&\,\,$0$&\,\,$0$&$(1,1)$&\,\,$2$&\,\,$0$&\,\,$0$&-$2$\\
&$+{b_{3}}+2\gamma$&$(1,1,2)$&\,\,$0$&\,\,$0$&-$2$&-$2$&-$2$&\,\,$0$&\,\,$0$&$(1,1)$&\,\,$2$&\,\,$0$&\,\,$0$&-$2$\\
\hline
\end{tabular}
\caption{Vector-like exotic states (continued).}
\end{table}
\end{center}

\section{Spectrum of Model in Section 6.2}
The following tables present the spectrum of the $\tilde{S}$ model given in Section \ref{subsec:Model2}.
As in Appendix A, all charges are multiplied by four and the CPT conjugates are omitted. Throughout the tables we use the vector combination:
$\zeta=1+b_1+b_2+b_3=\{\bar{\phi}^{1,...,8}\}$. We also use the following notation:
\begin{equation*}
Q_8= Q_{\overline{\phi}^{5}}+Q_{\overline{\phi}^{6}}+Q_{\overline{\phi}^{7}}+Q_{\overline{\phi}^{8}}~.\nonumber\\
\end{equation*}
\small{
\begin{table}[h]
\hspace*{-18 mm}
\renewcommand{\arraystretch}{1}
\begin{tabular}{| c | c | c | c | c c c c c c c c | c | c | c |}
    \hline

    F & Sector & Name & $(C,L)$ & $Q_C$ & $Q_L$ & $Q_{\bar{\eta}^1}$ & $Q_{\bar{\eta}^2}$ & $Q_{\bar{\eta}^3}$ & $Q_{\bar{y}^{3,6}}$ & $Q_{\bar{y}^1\bar{w}^5}$ & $Q_{\bar{w}^{2,4}}$ & $SU(2)_{1,...,4}$ & $SU(4)$&$Q_{8}$ \\ \hline

     b & NS & ($h$)                  & (1,2)   &~0 & -4 &~0 &~0 &~4  &~0  &~0  &~0  & (1,1,1,1) & ~1 &~0 \\
       &    & ($\bar{h}$)          & (1,$2$) &~0 &~4  &~0 &~0 & -4 &~0  &~0  &~0  & (1,1,1,1) & ~1 &~0 \\
       &    & $(\Phi_{56})$        & (1,1)   &~0 &~0  &~0 &~0 &~0  &~4  &~4  &~0  & (1,1,1,1) & ~1 &~0 \\
       &    & $(\bar{\Phi}_{56})$  & (1,1)   &~0 &~0  &~0 &~0 &~0  & -4 & -4 &~0  & (1,1,1,1) & ~1 &~0 \\
       &    & $(\Phi_{56}')$       & (1,1)   &~0 &~0  &~0 &~0 &~0  & -4 &~4  &~0  & (1,1,1,1) & ~1 &~0 \\
       &    & $(\bar{\Phi}_{56}')$ & (1,1)   &~0 &~0  &~0 &~0 &~0  &~4  & -4 &~0  & (1,1,1,1) & ~1 &~0 \\
       &    & $(\Phi_{46})$        & (1,1)   &~0 &~0  &~0 &~0 &~0  &~4  &~0  &~4  & (1,1,1,1) & ~1 &~0 \\
       &    & $(\bar{\Phi}_{46})$  & (1,1)   &~0 &~0  &~0 &~0 &~0  & -4 &~0  & -4 & (1,1,1,1) & ~1 &~0 \\
       &    & $(\Phi_{46}')$       & (1,1)   &~0 &~0  &~0 &~0 &~0  & -4 &~0  &~4  & (1,1,1,1) & ~1 &~0 \\
       &    & $(\bar{\Phi}_{46}')$ & (1,1)   &~0 &~0  &~0 &~0 &~0  &~4  &~0  & -4 & (1,1,1,1) & ~1 &~0 \\
       &    & $(\Phi_{45})$        & (1,1)   &~0 &~0  &~0 &~0 &~0  &~0  &~4  &~4  & (1,1,1,1) & ~1 &~0 \\
       &    & $(\bar{\Phi}_{45})$  & (1,1)   &~0 &~0  &~0 &~0 &~0  &~0  & -4 & -4 & (1,1,1,1) & ~1 &~0 \\
       &    & $(\Phi_{45}')$       & (1,1)   &~0 &~0  &~0 &~0 &~0  &~0  & -4 &~4  & (1,1,1,1) & ~1 &~0 \\
       &    & $(\bar{\Phi}_{45}')$ & (1,1)   &~0 &~0  &~0 &~0 &~0  &~0  &~4  & -4 & (1,1,1,1) & ~1 &~0 \\
       &    & $(\xi_{1,2,3})$      & (1,1)   &~0 &~0  &~0 &~0 &~0  &~0  &~0  &~0  & (1,1,1,1) & ~1 &~0 \\ \hline
  \end{tabular}
  \caption{The untwisted Neveu-Schwarz scalar states.}
\end{table}
}
\small{
\begin{table}[h]
\hspace*{-18 mm}
\renewcommand{\arraystretch}{1}
  \begin{tabular}{| c | c | c | c | c c c c c c c c | c | c | c |}
    \hline

    F & Sector & Name & $(C,L)$ & $Q_C$ & $Q_L$ & $Q_{\bar{\eta}^1}$ & $Q_{\bar{\eta}^2}$ & $Q_{\bar{\eta}^3}$ & $Q_{\bar{y}^{3,6}}$ & $Q_{\bar{y}^1\bar{w}^5}$ & $Q_{\bar{w}^{2,4}}$ & $SU(2)_{1,...,4}$ &  $SU(4)$ & $Q_{8}$ \\ \hline
     f & $\tilde{S} \text{ } ( +\zeta )$ & $\tilde{h}$& (1,2) &~0 &~4  &~0 &~0 &~0  &~0 &~0 &~0 &  (1,1,1,1) & ~4 &-4 \\
      
       &     &$\bar{\tilde{h}}$ & (1,$2$) &~0 & -4 &~0 &~0 &~0  &~0 &~0 &~0 &  (1,1,1,1) &~4  &~4 \\
       
       &     &$\xi_4$ & (1,1)         &~0 &~0  &~0 &~0 &~0  &~0 &~0 &~0 &  (1,2,2,1) &~1  &~0 \\
       &     &$\xi_5$ & (1,1)         &~0 &~0  &~0 &~0 &~0  &~0 &~0 &~0 &  (1,2,2,1) &~1  &~0 \\
       &     &$\xi_6$ & (1,1)         &~0 &~0  &~0 &~0 &~0  &~0 &~0 &~0 &  (2,1,1,2) &~1  &~0 \\
       &     &$\xi_7$ & (1,1)         &~0 &~0  &~0 &~0 &~0  &~0 &~0 &~0 &  (2,1,1,2) &~1  &~0 \\
       &     &$\phi_1$ & (1,1)         &~0 &~0  &~0 &~0 &~4  &~0 &~0 &~0 &  (1,1,1,1) & ~4 &-4 \\
       &     &$\phi_1$ & (1,1)         &~0 &~0  &~0 &~0 &-4  &~0 &~0 &~0 &  (1,1,1,1) & ~4 &~4 \\

    \hline
  \end{tabular}
  \caption{The $\tilde{S}$ and $\tilde{S}+\zeta$ sector.}
  \label{stildetab}
\end{table}
}

\small{
\begin{table}[h]
\hspace*{-18 mm}
\renewcommand{\arraystretch}{1}
  \begin{tabular}{| c | c | c | c | c c c c c c c c | c | c | c |}
    \hline

    F & Sector & Name & $(C,L)$ & $Q_C$ & $Q_L$ & $Q_{\bar{\eta}^1}$ & $Q_{\bar{\eta}^2}$ & $Q_{\bar{\eta}^3}$ & $Q_{\bar{y}^{3,6}}$ & $Q_{\bar{y}^1\bar{w}^5}$ & $Q_{\bar{w}^{2,4}}$ & $SU(2)_{1,...,4}$ &  $SU(4)$ & $Q_{8}$ \\ \hline

    f & $b_1$ & $Q_1$    & (3,2)         &~2  &~0  & -2 &~0 &~0 & -2 &~0 &~0 & (1,1,1,1) &~1 &~0 \\
      &       & $u_1^c$  & ($\bar{3}$,1) & -2 & -4 & -2 &~0 &~0 &~2  &~0 &~0 & (1,1,1,1) &~1 &~0 \\
      &       & $d_1^c$  & ($\bar{3}$,1) & -2 &~4  & -2 &~0 &~0 & -2 &~0 &~0 & (1,1,1,1) &~1 &~0 \\
      &       & $L_1$    & (1,2)         & -6 &~0  & -2 &~0 &~0 &~2  &~0 &~0 & (1,1,1,1) &~1 &~0 \\ 
      &       & $e_1^c$  & (1,1)         &~6  &~4  & -2 &~0 &~0 &~2  &~0 &~0 & (1,1,1,1) &~1 &~0 \\
      &       & $N_1^c$  & (1,1)         &~6  & -4 & -2 &~0 &~0 & -2 &~0 &~0 & (1,1,1,1) &~1 &~0 \\\hline

    f & $b_2$ & $Q_2$    & (3,2)         &~2  &~0  &~0 & -2 &~0 &~0 &~2  &~0 & (1,1,1,1) &~1 &~0 \\
\hline

  \end{tabular}
  \caption{The observable matter sectors.}
\label{nonSmapped2}
\end{table}
}
\small{
\begin{table}[h]
\setlength{\tabcolsep}{4.5 pt}
\hspace*{-18 mm}
\renewcommand{\arraystretch}{1}
  \begin{tabular}{| c | c | c | c | c c c c c c c c | c | c | c |}
    \hline

    F & Sector & Name&$(C,L)$ & $Q_C$ & $Q_L$ & $Q_{\bar{\eta}^1}$ & $Q_{\bar{\eta}^2}$ & $Q_{\bar{\eta}^3}$ & $Q_{\bar{y}^{3,6}}$ & $Q_{\bar{y}^1\bar{w}^5}$ & $Q_{\bar{w}^{2,4}}$ & $SU(2)_{1,...,4}$ & $SU(4)$ & $Q_{8}$ \\ \hline
        f & $\mathbf{1} + b_1 + b_2 + b_3 $      &$\zeta$& (1,1) &~0 &~0 &~0 &~0 &~0 & ~0 & ~0 &~0 & (1,1,1,1) &~4 &~0 \\
         &       &$\bar{\zeta}$& (1,1) &~0 &~0 &~0 &~0 &~0 & ~0 & ~0 &~0 & (1,1,1,1) &~4 &~0 \\\hline
    b & $b_1 + b_2$      &$\Phi^{\alpha\beta}_1$& (1,1) &~0 &~0 &~0 &~0 &~0 & ~2 & ~2 &~0 & (1,1,1,1) &~6 &~0 \\
     &  $+ \alpha+\beta \text{ } (+ \zeta)$     &$\bar{\Phi}^{\alpha\beta}_1$& (1,1) &~0 &~0 &~0 &~0 &~0 & -2 & -2 &~0 & (1,1,1,1) &~6 &~0 \\

      &                  &$\Phi^{\alpha\beta}_2$& (1,1) &~0 &~0 &~0 &~0 &~0 & -2 & ~2 &~0 & (1,1,2,2) &~1 &~0 \\
      &                  &$\bar{\Phi}^{\alpha\beta}_2$& (1,1) &~0 &~0 &~0 &~0 &~0 & ~2 & -2 &~0 & (1,1,2,2) &~1 &~0 \\
       &                  &$\Phi^{\alpha\beta}_3$& (1,1) &~0 &~0 &~0 &~0 &~0 & -2 & ~2 &~0 & (2,2,1,1) &~1 &~0 \\
\hline

  \end{tabular}
  \caption{The hidden sectors.}
  \label{hiddentable}
\end{table}
}

\small{
\begin{table}[h]
\setlength{\tabcolsep}{4 pt}
\hspace*{-16 mm}
\renewcommand{\arraystretch}{1}
  \begin{tabular}{| c | c | c | c | c c c c c c c c | c | c | c |}
    \hline

    F & Sector  & Name & $(C,L)$ & $Q_C$ & $Q_L$ & $Q_{\bar{\eta}^1}$ & $Q_{\bar{\eta}^2}$ & $Q_{\bar{\eta}^3}$ & $Q_{\bar{y}^{3,6}}$ & $Q_{\bar{y}^1\bar{w}^5}$ & $Q_{\bar{w}^{2,4}}$ & $SU(2)_{1,...,4}$ & $SU(4)$ & $Q_{8}$ \\ \hline

    b & $\alpha+\beta \text{ } (+\zeta) $  &$N_1$& (1,1) &~0 &~0 & -2 & ~2 &~0 &~0 &~0 &~0 & (1,1,1,1) & ~6 &~0 \\
      &                 &$\bar{N}_1$& (1,1) &~0 &~0 & ~2 & -2 &~0 &~0 &~0 &~0 & (1,1,1,1) & ~6 &~0 \\
      &                 &$N_2$& (1,1) &~0 &~0 & ~2 & -2 &~0 &~0 &~0 &~0 & (1,1,1,1) & ~6 &~0 \\
      &                 &$\bar{N}_2$& (1,1) &~0 &~0 & -2 & ~2 &~0 &~0 &~0 &~0 & (1,1,1,1) & ~6 &~0 \\ \hline

\end{tabular}
  \caption{$SO(10)$ singlets without $\tilde{S}$-partners.}
  \label{nonSmapped1}
\end{table}
}
\small{
\begin{table}[h]
\setlength{\tabcolsep}{4 pt}
\vspace{-16 mm}
\hspace*{-16 mm}
\renewcommand{\arraystretch}{1}
  \begin{tabular}{| c | c | c | c | c c c c c c c c | c | c | c |}
    \hline

    F & Sector &Name& $(C,L)$ & $Q_C$ & $Q_L$ & $Q_{\bar{\eta}^1}$ & $Q_{\bar{\eta}^2}$ & $Q_{\bar{\eta}^3}$ & $Q_{\bar{y}^{3,6}}$ & $Q_{\bar{y}^1\bar{w}^5}$ & $Q_{\bar{w}^{2,4}}$ & $SU(2)_{1,...,4}$ & $SU(4)$ & $Q_{8}$ \\ \hline

    f & $b_1 +~2\gamma \text{ } (+\zeta)$ &$V_1$& (1,1) &~0 &~0 &~0 & -2 &~2 & -2 &~0 &~0 & (1,1,2,2) & ~1 & ~0 \\
      &                 &$V_2$& (1,1) &~0 &~0 &~0 & -2 &~2 & -2 &~0 &~0 & (1,1,1,1) &~1  &-8  \\
      &                 &$V_3$& (1,1) &~0 &~0 &~0 & -2 &~2 & -2 &~0 &~0 & (1,1,1,1) &~1  &~8  \\
       
        &  &$V_{4}$& (1,1) &~0 &~0 &~0 & -2 &~2 & ~2 &~0 &~0 & (1,1,1,1) &~6 &~0 \\
      &                     &$V_{5}$& (1,1) &~0 &~0 &~0 & -2 &~2 & -2 &~0 &~0 & (2,2,1,1) &~1 &~0 \\ \hline
    f &$b_2+2\gamma\text{ } (+\zeta)$ &$V_6$& (1,1) &~0 &~0 & -2 &~0 &~2 &~0 & ~2 &~0 & (1,1,2,2) & ~1 & ~0 \\
      
      &                 &$V_7$& (1,1) &~0 &~0 & -2 &~0 &~2 &~0 & ~2 &~0 & (1,1,1,1) & ~1 &-8  \\
      &                 &$V_{8}$& (1,1) &~0 &~0 & -2 &~0 &~2 &~0 & ~2 &~0 & (1,1,1,1) &~1 & ~8 \\ 
       &  &$V_{9}$& (1,1) &~0 &~0 & -2 &~0 &~2 &~0 & -2 &~0 & (1,1,1,1) &~6 &~0 \\
      &                     &$V_{10}$& (1,1) &~0 &~0 & -2 &~0 &~2 &~0 & ~2 &~0 & (2,2,1,1) &~1 &~0 \\ \hline
     
    f & $b_3+2\gamma\text{ } (+\zeta)$ &$V_{11}$& (1,1) &~0 &~0 & -2 & -2 &~0 &~0 &~0 & -2 & (1,1,2,2) & ~1 & ~0 \\
      
      &                 &$V_{12}$& (1,1) &~0 &~0 & -2 & -2 &~0 &~0 &~0 &-2  & (1,1,1,1) & ~1 &~8 \\
      &                 &$V_{13}$& (1,1) &~0 &~0 & -2 & -2 &~0 &~0 &~0 &-2  & (1,1,1,1) & ~1 & -8 \\
       &  &$V_{14}$& (1,1) &~0 &~0 & -2 & -2 &~0 &~0 &~0 & ~2  & (1,1,1,1) &~6 &~0 \\
      &                     &$V_{15}$& (1,1) &~0 &~0 & -2 & -2 &~0 &~0 &~0 & -2  & (2,2,1,1) &~1 &~0 \\ \hline

\end{tabular}
  \caption{$SO(10)$ singlets with $\tilde{S}$-partners.}
\label{so10sinws}
\end{table}
}

\small{
\begin{table}[h]
\setlength{\tabcolsep}{4 pt}
\vspace{-16 mm}
\hspace*{-16 mm}
\renewcommand{\arraystretch}{1}
  \begin{tabular}{| c | c | c | c | c c c c c c c c | c | c | c |}
    \hline

    F & Sector &Name& $(C,L)$ & $Q_C$ & $Q_L$ & $Q_{\bar{\eta}^1}$ & $Q_{\bar{\eta}^2}$ & $Q_{\bar{\eta}^3}$ & $Q_{\bar{y}^{3,6}}$ & $Q_{\bar{y}^1\bar{w}^5}$ & $Q_{\bar{w}^{2,4}}$ & $SU(2)_{1,...,4}$ & $SU(4)$ & $Q_{8}$ \\ \hline

    b & $\tilde{S}+b_1+2\gamma \text{ } (+\zeta)$ &$V_{16}$& (1,1) &~0 &~0  &~0 &~2 &~2  & -2 &~0 &~0 & (1,1,1,1) &~4 & -4 \\

      &                         &$V_{17}$& (1,1) &~0 &~0  &~0 &~2 &~2  & ~2 &~0 &~0 & (1,1,1,1) &~4 &~4  \\
      &                         &$V_{18}$& (1,1) &~0 &~0  &~0 &~2 & -2 & ~2 &~0 &~0 & (1,2,1,2) &~1 &~0  \\
      &     &$V_{19}$& (1,1) &~0 &~0 &~0 & ~2 & -2 & ~2 &~0 &~0 & (2,1,2,1) & ~1 &~0 \\
       \hline
      
    b & $\tilde{S} + b_2 + 2\gamma \text{ } (+\zeta)$  &$V_{20}$& (1,1) &~0 &~0 & ~2 &~0 & ~2 &~0 & ~2  &~0 & (1,1,1,1) &~4 & -4 \\
     
      &                             &$V_{21}$& (1,1) &~0 &~0 & ~2 &~0 & ~2 &~0 & -2  &~0 & (1,1,1,1) &~4 & ~4 \\ 
      
      &              &$V_{22}$& (1,1) &~0 &~0 & ~2 &~0 & -2 &~0 & -2  &~0 & (1,2,1,2) &~1 &~0  \\

      &     &$V_{23}$& (1,1) &~0 &~0 & ~2 &~0 & -2 &~0 & -2 &~0 & (2,1,2,1) & ~1 &~0 \\ \hline
    
    b & $\tilde{S} + b_3 +~2\gamma \text{ } (+\zeta)$ &$V_{24}$& (1,1) &~0 &~0 & ~2 & -2 &~0 &~0 &~0 & ~2 & (1,1,1,1) &~4 &~4  \\
     
      &                             &$V_{25}$& (1,1) &~0 &~0 & ~2 & -2 &~0 &~0 &~0 & -2 & (1,1,1,1) &~4 & -4 \\
      
      &                             &$V_{26}$& (1,1) &~0 &~0 & ~2 &~2  &~0 &~0 &~0 & ~2 & (1,2,1,2) &~1 &~0  \\ 
    
      &    &$V_{27}$& (1,1) &~0 &~0 & ~2 & ~2 &~0 &~0 &~0 & ~2 & (2,1,2,1) & ~1 &~0 \\ \hline

\end{tabular}
  \caption{$SO(10)$ singlets' $\tilde{S}$-partners.}
\label{so10sinws1}
\end{table}
}

\small{
\begin{table}[h]
\setlength{\tabcolsep}{4 pt}
\hspace*{-16 mm}
\renewcommand{\arraystretch}{1}
  \begin{tabular}{| c | c | c | c | c c c c c c c c | c | c | c |}
    \hline

    F & Sector & Name & $(C,L)$ & $Q_C$ & $Q_L$ & $Q_{\bar{\eta}^1}$ & $Q_{\bar{\eta}^2}$ & $Q_{\bar{\eta}^3}$ & $Q_{\bar{y}^{3,6}}$ & $Q_{\bar{y}^1\bar{w}^5}$ & $Q_{\bar{w}^{2,4}}$ & $SU(2)_{1,...,4}$ & $SU(4)$ & $Q_{8}$ \\ \hline

    f & $b_2 + \beta$  &$H_1$& (1,2) &~0 &~0 &~0 &~0 &~0 & ~2 &~0 & -2  & (1,2,1,1) &~1 &~0 \\
      &                &$\bar{H}_1$& (1,2) &~0 &~0 &~0 &~0 &~0 & -2 &~0 & ~2  & (1,2,1,1) &~1 &~0 \\ \hline

    b & $\tilde{S} + b_2 + \beta$ &$H_2$& (1,2) &~0 &~0 &~0 &~0 &~0 & ~2 &~0 & -2  & (1,1,2,1) &~1 &~0 \\
      &                           &$\bar{H}_2$& (1,2) &~0 &~0 &~0 &~0 &~0 & -2 &~0 & ~2  & (1,1,2,1) &~1 &~0 \\ \hline

    b & $b_2 + b_3 $        &$H_3$& (1,1) & -3 & ~2 & ~1 & ~1 & -1 & -2 &~0 &~0 & (1,2,1,1) & ~1 & ~4 \\
      & $+\beta \pm \gamma  $ &$\bar{H}_3$& (1,1) & ~3 & -2 & -1 & -1 & ~1 & ~2 &~0 &~0 & (1,2,1,1) & ~1 & -4 \\ 
          &  &$H_{4}$& (1,1) & -3 & ~2 & ~1 & ~1 & -1 & -2 &~0 &~0 & (2,1,1,1) & ~1 & -4 \\
\hline

\end{tabular}
  \caption{Exotic states with $\tilde{S}$-partners (i).}
\label{exotics1}
\end{table}
}

\small{
\begin{table}[h]
\setlength{\tabcolsep}{4 pt}
\hspace*{-16 mm}
\renewcommand{\arraystretch}{1}
  \begin{tabular}{| c | c | c | c | c c c c c c c c | c | c | c |}
    \hline

    F & Sector & Name & $(C,L)$ & $Q_C$ & $Q_L$ & $Q_{\bar{\eta}^1}$ & $Q_{\bar{\eta}^2}$ & $Q_{\bar{\eta}^3}$ & $Q_{\bar{y}^{3,6}}$ & $Q_{\bar{y}^1\bar{w}^5}$ & $Q_{\bar{w}^{2,4}}$ & $SU(2)_{1,...,4}$ & $SU(4)$ & $Q_{8}$ \\ \hline

    b & $b_1+b_2$ &$H_{17}$& ($\bar{3}$,1)         &-1  & ~2 & ~1 & ~1 & ~1 &-2 &-2 & ~0 & (1,1,1,1) &~1  &-4  \\
      &   $+\alpha +\beta \pm \gamma + \zeta$                &$\bar{H}_{17}$& ($3$,1) & ~1 & -2 & -1 & -1 & -1 &~2 &~2 &~0  & (1,1,1,1) & ~1 & ~4 \\
      &                    &$H_{18}$& (1,2)         & ~3 & -2 & ~1 & ~1 & ~1 &-2 &-2 & ~0 & (1,1,1,1) &~1  &-4  \\
      &                    &$\bar{H}_{18}$& (1,$2$) & -3 & ~2 & -1 & -1 & -1 &~2 &~2 &~0  & (1,1,1,1) & ~1 & ~4 \\
      &                    &$H_{19}$& (1,1)         & ~3 & ~2 & -3 & ~1 & -1 &-2 &-2 & ~0 & (1,1,1,1) &~1  &-4  \\
      &                    &$\bar{H}_{19}$& (1,1)         & -3 & -2 &~3  & -1 & ~1 &~2 &~2 &~0  & (1,1,1,1) & ~1 & ~4 \\
      &                    &$H_{20}$& (1,1)         & ~3 & ~2 & ~1 & -3 & -1 &-2 &-2 & ~0 & (1,1,1,1) &~1  &-4  \\
\hline

      \end{tabular}
  \caption{Exotic states with $\tilde{S}$-partners (ii).}
\label{exotics2}
\end{table}
}

\small{
\begin{table}[h]
\setlength{\tabcolsep}{4 pt}
\hspace*{-16 mm}
  \begin{tabular}{| c | c | c | c | c c c c c c c c | c | c | c |}
    \hline

    F & Sector & Name & $(C,L)$ & $Q_C$ & $Q_L$ & $Q_{\bar{\eta}^1}$ & $Q_{\bar{\eta}^2}$ & $Q_{\bar{\eta}^3}$ & $Q_{\bar{y}^{3,6}}$ & $Q_{\bar{y}^1\bar{w}^5}$ & $Q_{\bar{w}^{2,4}}$ & $SU(2)_{1,...,4}$ & $SU(4)$ & $Q_{8}$ \\ \hline
      
    f & $b_1 + \alpha$ &$H_{35}$& (1,$2$) &~0 &~0 &~0 &~0 &~0 &~0 & -2 & ~2 & (2,1,1,1) &~1 &~0 \\
     
      &                &$\bar{H}_{35}$& (1,$2$) &~0 &~0 &~0 &~0 &~0 &~0 & ~2 & -2 & (2,1,1,1) &~1 &~0 \\
      \hline
        
    b & $\tilde{S}+b_1+b_2+b_3$  &$H_{36}$& (1,1) &~0 & ~4 &~0 & -2 & ~2 &~0 &~0 &~0 & (1,1,1,2) &~1 & ~0 \\
    &       $+\alpha+\zeta$     &$\bar{H}_{36}$& (1,1) &~0 & -4 &~0 & ~2 & -2 &~0 &~0 &~0 & (1,1,1,2) &~1 & ~0 \\
    &                       &$H_{37}$& (1,1) &~0 & ~4 &~0 & -2 & ~2 &~0 &~0 &~0 & (1,1,1,2) &~1 & ~0 \\
    &                       &$\bar{H}_{37}$& (1,1) &~0 & -4 &~0 & ~2 & -2 &~0 &~0 &~0 & (1,1,1,2) &~1 & ~0 \\

       \hline

    b & $\tilde{S}+b_1+b_2$ &$H_{38}$& (1,2) &~0 &~0  & -2 &~0 & -2 &~0 &~0 &~0 & (1,1,2,1) &~1 &~0 \\
\hline

      \end{tabular}
  \caption{Exotic states without $\tilde{S}$-partners.}
  \label{nonSmapped4}
\end{table}
}


\newpage
\bibliographystyle{unsrturl}
\bibliography{Reference}

\end{document}